\documentclass[twocolumn,twocolappendix]{aastex63}
\NewPageAfterKeywords

\newcommand{\simprop}{\mathrel{\vcenter{
  \offinterlineskip\halign{\hfil$##$\cr
    \propto\cr\noalign{\kern2pt}\sim\cr\noalign{\kern-2pt}}}}}
    
\graphicspath{{./}{figures/}}

\begin{document}
\title{Multiband imaging of the HD 36546 debris disk: a refined view from SCExAO/CHARIS\footnote{Based on data collected at Subaru Telescope, which is operated by the National Astronomical Observatory of Japan.}}

%Authors:
\correspondingauthor{Kellen Lawson}
\email{kellenlawson@gmail.com}
\author{Kellen Lawson}
\affiliation{Department of Physics and Astronomy, University of Oklahoma, Norman, OK}

\author{Thayne Currie}
\affiliation{Subaru Telescope, National Astronomical Observatory of Japan, 
650 North A`oh$\bar{o}$k$\bar{u}$ Place, Hilo, HI  96720, USA}
\affiliation{NASA-Ames Research Center, Moffett Blvd., Moffett Field, CA, USA}
\affiliation{Eureka Scientific, 2452 Delmer Street Suite 100, Oakland, CA, USA}

\author{John P. Wisniewski}
\affiliation{Department of Physics and Astronomy, University of Oklahoma, Norman, OK}

\author{Motohide Tamura}
\affil{Astrobiology Center of NINS, 2-21-1, Osawa, Mitaka, Tokyo, 181-8588, Japan}
\affiliation{Department of Astronomy, Graduate School of Science, The University of Tokyo, 7-3-1, Hongo, Bunkyo-ku, Tokyo, 113-0033, Japan}
\affiliation{National Astronomical Observatory of Japan, 2-21-2, Osawa, Mitaka, Tokyo 181-8588, Japan}

\author{Jean-Charles Augereau}
\affil{Univ. Grenoble Alpes, CNRS, IPAG, 38000 Grenoble, France}

\author{Timothy D. Brandt}
\affiliation{Department of Physics, University of California, Santa Barbara, Santa Barbara, California, USA}

\author{Olivier Guyon}
\affiliation{Subaru Telescope, National Astronomical Observatory of Japan, 
650 North A`oh$\bar{o}$k$\bar{u}$ Place, Hilo, HI  96720, USA}
\affil{Steward Observatory, The University of Arizona, Tucson, AZ 85721, USA}
\affil{The Wyant College of Optical Sciences, University of Arizona, Tucson, AZ 85721, USA}
\affil{Astrobiology Center of NINS, 2-21-1, Osawa, Mitaka, Tokyo, 181-8588, Japan}

\author{N. Jeremy Kasdin}
\affiliation{College of Arts and Sciences, University of San Francisco, San Francisco, CA, USA}
\affiliation{Department of Mechanical Engineering, Princeton University, Princeton, NJ, USA}

\author{Tyler D. Groff}
\affiliation{NASA-Goddard Space Flight Center, Greenbelt, MD, USA}

\author{Julien Lozi}
\affiliation{Subaru Telescope, National Astronomical Observatory of Japan, 
650 North A`oh$\bar{o}$k$\bar{u}$ Place, Hilo, HI  96720, USA}

\author{Vincent Deo}
\affiliation{Subaru Telescope, National Astronomical Observatory of Japan, 
650 North A`oh$\bar{o}$k$\bar{u}$ Place, Hilo, HI  96720, USA}

\author{Sebastien Vievard}
\affiliation{Subaru Telescope, National Astronomical Observatory of Japan, 
650 North A`oh$\bar{o}$k$\bar{u}$ Place, Hilo, HI  96720, USA}

\author{Jeffrey Chilcote}
\affiliation{Department of Physics, University of Notre Dame, South Bend, IN, USA}

\author{Nemanja Jovanovic}
\affiliation{Department of Astronomy, California Institute of Technology, 1200 East California Boulevard, Pasadena, CA 91125}
\author{Frantz Martinache}
\affiliation{Universit\'{e} C\^{o}te d'Azur, Observatoire de la C\^{o}te d'Azur, CNRS, Laboratoire Lagrange, France}
\author{Nour Skaf}
\affiliation{Subaru Telescope, National Astronomical Observatory of Japan, 
650 North A`oh$\bar{o}$k$\bar{u}$ Place, Hilo, HI  96720, USA}

\affil{LESIA, Observatoire de Paris, Université PSL, CNRS, Sorbonne Universit\'e, Universit\'e de Paris, 5 place Jules Janssen, 92195 Meudon, France}
	\affil{Department of Physics and Astronomy, University College London, London, United Kingdom}

\author{Thomas Henning}
\affiliation{Max Planck Institute for Astronomy, Königstuhl 17, D-69117 Heidelberg, Germany}

\author{Gillian Knapp}
\affiliation{Department of Astrophysical Science, Princeton University, Peyton Hall, Ivy Lane, Princeton, NJ 08544, USA}

\author{Jungmi Kwon}
\affiliation{Department of Astronomy, Graduate School of Science, The University of Tokyo, 7-3-1, Hongo, Bunkyo-ku, Tokyo, 113-0033, Japan}

\author{Michael W. McElwain}
\affiliation{NASA-Goddard Space Flight Center, Greenbelt, MD, USA}

\author{Tae-Soo Pyo}
\affiliation{Subaru Telescope, National Astronomical Observatory of Japan, 
650 North A`oh$\bar{o}$k$\bar{u}$ Place, Hilo, HI  96720, USA}

\author{Michael L. Sitko}
\affiliation{Space Science Institute, 475 Walnut Street, Suite 205, Boulder, CO 80301, USA}

\author{Taichi Uyama}
\affiliation{Infrared Processing and Analysis Center, California Institute of Technology, 1200 E. California Boulevard, Pasadena, CA 91125, USA}
\affiliation{NASA Exoplanet Science Institute, Pasadena, CA 91125, USA}
\affiliation{National Astronomical Observatory of Japan, 2-21-2, Osawa, Mitaka, Tokyo 181-8588, Japan}

\author{Kevin Wagner}
\affiliation{NASA Hubble Fellowship Program - Sagan Fellow}
\affiliation{Steward Observatory, The University of Arizona, Tucson, AZ 85721, USA}

\submitjournal{AJ}
\accepted{2021 September 17}

\begin{abstract}
We present the first multi-wavelength (near-infrared; $1.1 - 2.4$ $\micron$) imaging of HD 36546's debris disk, using the Subaru Coronagraphic Extreme Adaptive Optics (SCExAO) system coupled with the Coronagraphic High Angular Resolution Imaging Spectrograph (CHARIS). As a 3-10 Myr old star, HD 36546 presents a rare opportunity to study a debris disk at very early stages. SCExAO/CHARIS imagery resolves the disk over angular separations of $\rho \sim 0\farcs25 - 1\farcs0$ (projected separations of $\rm{r_{proj}} \sim 25 - 101$ $\rm{au}$) and enables the first spectrophotometric analysis of the disk. The disk's brightness appears symmetric between its eastern and western extents and it exhibits slightly blue near-infrared colors on average (e.g. \textit{J}$-$\textit{K}$=-0.4\pm0.1$) -- suggesting copious sub-micron sized or highly porous grains. Through detailed modeling adopting a Hong scattering phase function (SPF), instead of the more common Henyey-Greenstein function, and using the differential evolution optimization algorithm, we provide an updated schematic of HD 36546's disk. The disk has a shallow radial dust density profile ($\alpha_{in} \approx 1.0$ and $\alpha_{out} \approx -1.5$), a fiducial radius of $r_0 \approx 82.7$ au, an inclination of $i \approx 79\fdg1$, and a position angle of \textit{PA} $ \approx 80\fdg1$. Through spine tracing, we find a spine that is consistent with our modeling, but also with a ``swept-back wing'' geometry. Finally, we provide constraints on companions, including limiting a companion responsible for a marginal \textit{Hipparcos}--\textit{Gaia} acceleration to a projected separation of $\lesssim 0\farcs2$ and to a minimum mass of $\lesssim 11$ $\rm M_{Jup}$.
\end{abstract}


\section{Introduction} \label{sec:intro}
    Gas depleted, dust-dominated \textit{debris disks} are critical targets for understanding the composition, structure, and (typically late-stage) formation of planetary systems \citep{Wyatt2008, Kenyon2008,Hughes2018}.  Scattered-light imaging of debris disks can reveal and characterize signatures of planets responsible for shaping the disks' morphologies and help constrain the properties of the planets themselves \citep[e.g.][]{Kalas2005,Lagrange2010}. ``Extreme adaptive-optics'' (exAO) facilities, such as SPHERE \citep{Beuzit2019}, GPI \citep{Macintosh2015}, and SCExAO \citep{Jovanovic2015,Lozi2018,Currie2020spie}, now deliver high-fidelity images of numerous debris disks at sub-arcsecond separations \citep[e.g.][]{Milli2017,Duchene2020,Esposito2020,Lawson2020}.
    
    Modeling of multi-wavelength imaging can enable analysis of the properties and composition of the constituent dust \citep[e.g.][]{Boccaletti2003,Fitzgerald2007,Goebel2018,Lawson2020}, providing a context for the chemistry and evolution of our own Kuiper belt \citep{Currie2015b,Milli2017,Chen2020}. However, the assumed scattering phase function (SPF) critically affects resulting model disk images and thus our ability to infer disk properties. The traditionally adopted Henyey-Greenstein (H-G) SPF \citep{Henyey1941} can reproduce scattered light images at large scattering angles but is not physically motivated and is often inconsistent with data at smaller scattering angles \citep[e.g.][]{Hughes2018}. Combined with evidence suggesting that the best-fit H-G SPF from disk modeling is correlated with the scattering angles probed by the data (rather than any more meaningful physical properties of the system), \citet{Hughes2018} suggest that some other SPF -- capable of explaining debris disk imagery irrespective of the scattering angles probed -- is needed. \citet{Lawson2020} showed that the \citet{Hong1985} SPF, which is derived from observations of solar system zodiacal dust, reasonably reproduced near-infrared (NIR) imagery of the debris disk of HD 15115. This finding is consistent with the suggestion in \citet{Hughes2018} that a near-universal SPF may exist to explain observed scattered light of a particular wavelength originating from nearly any circumstellar dust. Moreover, \citet{Lawson2020} found that the adopted SPF can fundamentally change the interpretation of scattered light images, e.g. the number of separate dust populations required to reproduce the data. If scattered light imagery of additional systems can be well-modeled using this SPF, it could provide further evidence of such a universal SPF.
    
    In this study, we present the first multi-wavelength scattered light imaging of the debris disk around HD 36546 \citep{Currie2017}, using the Subaru Coronagraphic Extreme Adaptive Optics (SCExAO) system and the Coronagraphic High Angular Resolution Imaging Spectrograph (CHARIS) integral field spectrograph in broadband (spanning NIR $J$, $H$, and $K$ bands, $1.13-2.39$ $\micron$) mode \citep{Groff2016}.   HD 36546's disk is highly inclined, bright in scattered light, and has an extremely high fractional luminosity ($L_{\rm IR}/L_{\star}$ $\sim$ 4$\times$10$^{-3}$) comparable to long-studied debris disks around $\beta$ Pictoris and HR 4796A and new benchmarks such as HD 115600 \citep{Smith1984,Schneider2014,Currie2015b,Millar-Blanchaer2015,Gibbs2019}.   
    
    The HD 36546 system lies at a distance of $101.35\pm0.73$ pc \citep{Gaia2018}, slightly foreground to the Taurus-Auriga star-forming region and is a member of the 3--10 Myr old 118 Tau association (99.1\% probability, \citealt{Gagne2018})\footnote{This association is labeled as ``Mamajek 17" in \citet{Currie2017}.}.  Due to the system's extreme youth, the HD 36546's debris disk represents a rare opportunity to study debris disk structure and composition at very early stages where protoplanetary disk accretion ceases, colder gas dissipates, and collisions between icy planetesimals tracing Kuiper belt-analogue formation are predicted to yield detectable debris at peak luminosity \citep{Currie2008,Currie2009,Cloutier2014,Ribas2015,Kenyon2008}.
    
    Previous work on HD 36546 found evidence for a disk with an extremely large scale height and strong forward-scattering: an H-G $g$ value of $\sim$0.7--0.85, among the highest of any known debris disk \citep{Currie2017,Hughes2018}.  Revisiting HD 36546's disk with our new, comparable-quality data obtained over a wider wavelength range will provide improved constraints on its structure and the first look at its composition.  Much like for the case with HD 15115 \citep{Lawson2020}, modeling these data with a different SPF may lead to revisions in our understanding of the HD 36546 disk's properties. 
    
    Our new data resolve the disk down to separations as small as $\sim0\farcs25$ ($\sim 25$ au) and enable the first spectrophotometric analysis of the disk. From these data, we provide surface brightness, surface contrast, and asymmetry profiles in CHARIS broadband and $J$, $H$, and $K$ bands -- as well as disk color profiles in $\rm J-H $ and $\rm H-K$. Through detailed forward modeling, we provide a refined schematic for HD 36546's disk. Finally, we provide constraints for the presence of companions within the CHARIS field of view (FOV; angular separations of $\rm 0\farcs11 \lesssim \rho \lesssim 1\farcs0$).

\section{Data}\label{sec:data}
    \subsection{Observations}
    We observed HD 36546 \citep[A0--A2V, V=6.95, H=6.92;][]{Currie2017, Lisse2017} on 2019 January 12 using the Subaru Telescope’s SCExAO running at 2 kHz paired with the CHARIS integral field spectrograph \citep{Groff2016} operating in low-resolution (R $\sim 20$), broadband (1.16–2.37 $\mu m$) mode, and utilizing SCExAO's Lyot coronagraph with 217 mas diameter occulting spot. Conditions were photometric with good 0\farcs{}4--0\farcs{}5 seeing and 16 km/hour winds\footnote{However, AO performance was compromised compared to typical performance due to a significant (4--5x) loss in the IR secondary mirror's reflectivity at blue optical wavelengths where AO188 (which provides a needed low-order correction for SCExAO) does wavefront sensing.   This loss resulted from oxidization of the mirror's silver coating from sulfur dioxide stemming from the Kilauea eruption in May 2018.   For wavefront sensing, HD 36546 was effectively a magnitude 10.5 star.   The mirrors were realuminized with a fresh coating in late 2019.}.  
    The data were collected in \textit{angular differential imaging} mode (ADI; \citealt{Marois2006}) and achieved total parallactic angle rotation of $99^{\circ}$ with total integration time of $t_{int} =$ 50 minutes. The data set is made up of 66 individual exposures, with 65 having exposure times of 45.73 seconds and one having an exposure time of 30.98 seconds. Sky frames were also obtained following the science data to enable sky subtraction.   
    
    A second target, HR 2466 (A2V; V = 5.21, H = 5.07), was observed on the same night with the same instrument configuration to enable \textit{reference star differential imaging} (RDI). These data include 143 individual exposures of 20.65 seconds each, for a total integration time of 49 minutes. For all reduction and analysis, we adopt the revised CHARIS pixel scale of 0\farcs0162 and the North-up PA offset of 2\fdg2 as reported in \citet{Currie2018}.

    \subsection{CHARIS Data Reduction}\label{sec:reduc}
    CHARIS data cubes were extracted from raw CHARIS exposures using the CHARIS Data Reduction Pipeline \citep{Brandt2017}, using modifications to read noise suppression listed in \citet{Currie2020spie}.  Extracted data take the form of image cubes with dimensions $(N_\lambda, N_x, N_y) = (22,201,201)$ (i.e. $201 \times 201$ pixel images for each of 22 wavelength channels). Subsequent basic image processing -- e.g. sky subtraction, image registration, spectrophotometric calibration --  was carried out as in \citet{Currie2011, Currie2018}.   Inspection of registered data cubes showed that the quality of the AO correction varied between exposures taken before (good), during (worst\footnote{This is a result of HD 36546 transiting very near zenith from Subaru, making alt/az tracking through transit challenging.}), and after (best) transit. Particularly low quality frames were identified by first measuring the peak-to-halo flux ratio of the satellite spots in each exposure\footnote{The peak flux is defined as the flux within an aperture with a radius of one instrumental full-width at half-maximum (FWHM), centered on the satellite spot. The halo flux is the flux within an annulus centered on the satellite spot, with inner and outer radii of $2\cdot$FWHM$-dr/2$ and $2\cdot$FWHM$+dr/2$ respectively, where dr is the radial extent that results in an annulus enclosing the same area as the peak aperture.}. These values were then averaged over the 4 satellite spots and 22 wavelength channels, and the median and median absolute deviation of the values across the full data sequence were computed. Any exposures falling more than 3 median absolute deviations below the median peak-to-halo ratio were removed. We removed three frames near transit which met this criteria.

    PSF subtraction to recover HD 36546's disk was performed using RDI with the \textit{Karhunen-Lo\`{e}ve Image Projection} (KLIP; \citealt{Soummer2012}) algorithm (RDI-KLIP), as well as ADI with both KLIP (ADI-KLIP) and the \textit{Adaptive, Locally Optimized Combination of Images} (A-LOCI; \citealt{Currie2012,Currie2015}) algorithm (ADI-ALOCI) {which is based on the LOCI algorithm \citep{Lafreniere2007}}. To search for planets, we exploit both ADI and \textit{spectral differential imaging} \citep[SDI;][]{SparksFord2002} \citep[see e.g.,][]{Beuzit2019}, using a more aggressive implementation of A-LOCI \citep[as in e.g.,][]{Currie2018,Currie2020spie,Currie2020b,Lawson2020}.  
    Unlike prior implementations, we performed an ADI-based reduction on the post-SDI residuals instead of SDI on the post-ADI residuals.   We found that for data sets with somewhat variable AO corrections leading to temporally decorrelated images, the former approach is slightly superior.   The utilized algorithm parameters are provided in Table \ref{tab:psfsub_settings}. 

    \begin{deluxetable*}{@{\extracolsep{0pt}}ccccccc}
    \tablewidth{0pt}
    \tablecaption{PSF Subtraction Algorithm Settings}
    \tablehead{
    \colhead{Method} & \colhead{Param. Tuning} & \colhead{Radial Prof. Sub.} & \colhead{$r_{min}$} & \colhead{$r_{max}$} & \colhead{$\Delta r_{sub}$} & \colhead{Algorithm Parameters}
    }
    \startdata
    RDI-KLIP & disk & False & 15 & 70 & 55 & $\rm{N_{PCA}} = 10$, $\rm{N_{zones}} = 1$,  \\
    ADI-KLIP & disk & True & 15 & 65 & 50 & $\rm{N_{PCA}} = 15$, $\rm{N_{zones}} = 1$, $\delta_{\rm{FWHM}} = 2.5$ \\
    ADI-ALOCI & disk & True & 15 & 70 & 5 & $\rm{N_{A}} = 800$, $g = 0.1$, $\delta_{\rm{FWHM}} = 2.5$, $\rm{SVD}=0.02$ \\
    SDI+ADI-ALOCI & planet & True & 5 & 70 & 2.5 & $\rm{N_{A}} = 50$, $g = 0.25$, $\delta_{\rm{FWHM}} = 0.55$, $\rm{SVD}=1.05\times10^{-6}$ \\
    \enddata
    \tablecomments{Algorithm settings for PSF subtraction with each of the three techniques utilized. `Radial Profile Sub.' indicates whether each reduction was carried out on radial profile subtracted data. $r_{min}$ and $r_{max}$ refer to the inner and outer radii of the subtraction region for each reduction (in pixels). $\Delta r_{sub}$ indicates the radial width of individual subtraction annuli. `g' refers to the aspect ratio of the optimization regions (which are annular subsections, defined in polar-coordinates by radial and azimuthal extents) with $g < 1$ producing azimuthally elongated regions and $g > 1$ producing radially elongated regions. `$N_A$' refers to the area of optimization regions in units of PSF cores. `$\delta_{\rm{FWHM}}$' indicates the minimum rotation gap in units of PSF FWHM (for both A-LOCI and KLIP). `$\Delta r_{\rm{sub}}$' gives the radial size of subtraction regions in units of pixels (for both A-LOCI and KLIP). `$N_{\rm{PCA}}$' indicates the number of principal components utilized in construction of the model PSF. `$N_{\rm{zones}}$' is the number of subsections into which each KLIP optimization annulus was divided (with a value of 1 corresponding to full annuli). `SVD' refers to the threshold for normalized singular values below which A-LOCI coefficients are truncated. For the ADI-ALOCI reduction, the algorithm is not restricted in the number of images that may be used in constructing the PSF model.  For the SDI+ADI-ALOCI parameter tuning, we also employed a moving pixel mask over the subtraction zone \citep{Marois2010b,Currie2012} and restricted the reference PSF construction for each target image to the 50 best-correlated frames \citep{Currie2012b}.
    }\label{tab:psfsub_settings}
    \end{deluxetable*}

    \subsection{Results}\label{sec:results}
\begin{figure*}%reducs.pdf
    \includegraphics[width=\textwidth]{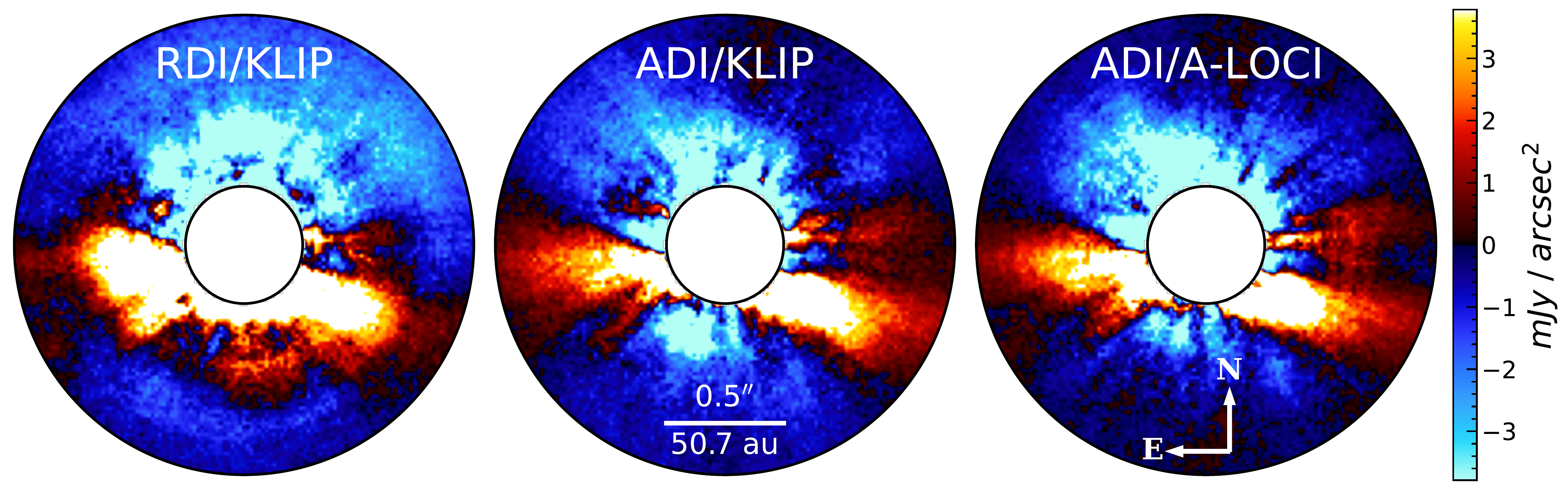}
    \caption{Wavelength-collapsed final images for all three reduction methods noted in Section \ref{sec:reduc}. The inner and outer masks in each panel have radii of 0\farcs25 and 1\farcs0 respectively.
    \label{fig:reducs}
    }
    \end{figure*}

\begin{figure*}%log_reducs.pdf
    \includegraphics[width=\textwidth]{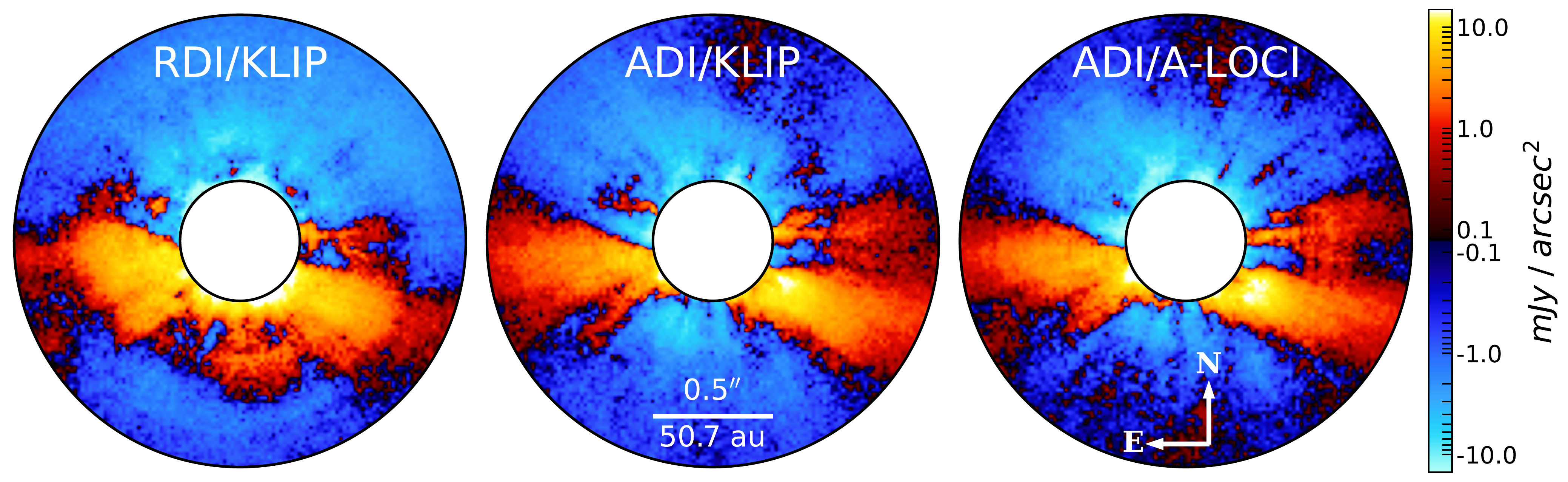}
    \caption{As Figure \ref{fig:reducs}, but with a symmetric logarithmic color scale. This scaling is linear in the range [-0.1, 0.1], and logarithmic otherwise.
    \label{fig:log_reducs}
    }
    \end{figure*}

    In broadband (wavelength-collapsed) imagery, all three PSF-subtraction techniques result in clear detection of the bright side of HD 36546's disk to separations as small as $\sim 0\farcs25$ (Figures \ref{fig:reducs}, \ref{fig:log_reducs}). Each result also shows a partial, weak detection of the fainter side of the disk in the west, which is strongest in the ADI-ALOCI result, with signal-to-noise ratio per resolution element (SNRE) $\sim 2.5$. As in the discovery imagery of the disk from SCExAO/HiCIAO \citep{Currie2017}, SCExAO/CHARIS imagery of HD 36546 reveals a highly inclined, significantly asymmetrically scattering disk. ADI-based reductions reveal no obvious evidence of ansae, indicative of gaps or a ring-like structure, within the field of view. The RDI-KLIP reduction, however, appears to show a change in the disk profile at $\rho \sim 0\farcs75$, consistent with the fiducial radius of the best-fitting disk model identified in \citet{Currie2017}.

\begin{figure*}%snre.pdf
    \includegraphics[width=\textwidth]{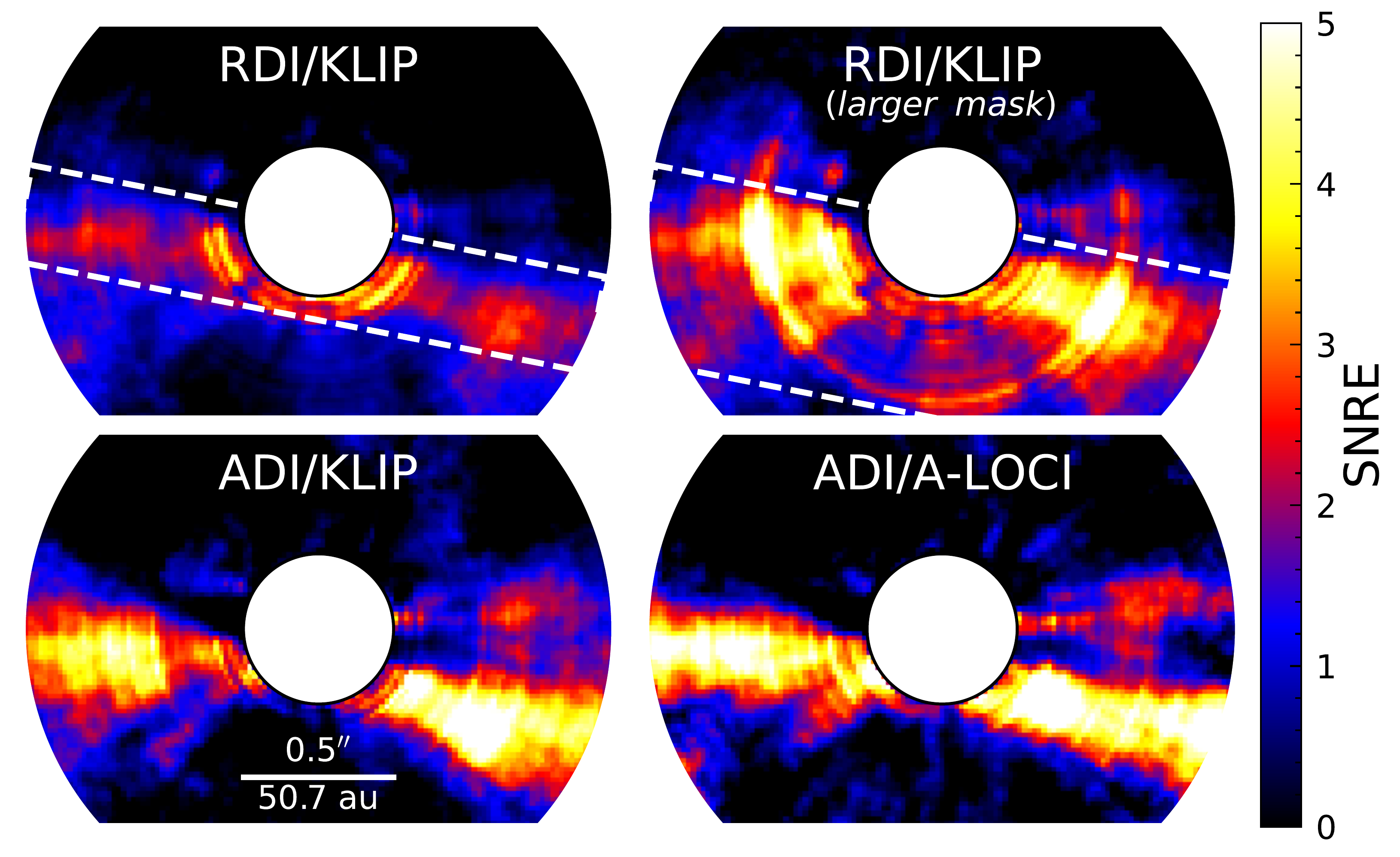}
    \caption{Signal-to-noise ratio per resolution element for the reductions shown in Figure \ref{fig:reducs}. For all panels, a mask is applied over the disk when the noise is calculated. The RDI reduction is shown using two different mask sizes (overlaid dashed white boxes), as the chosen size was found to have a strong impact on the resulting SNRE. Both ADI reductions use the smaller mask.
    \label{fig:snre}
    }
    \end{figure*}

\begin{figure*}%log_jhk.pdf
    \includegraphics[width=\textwidth]{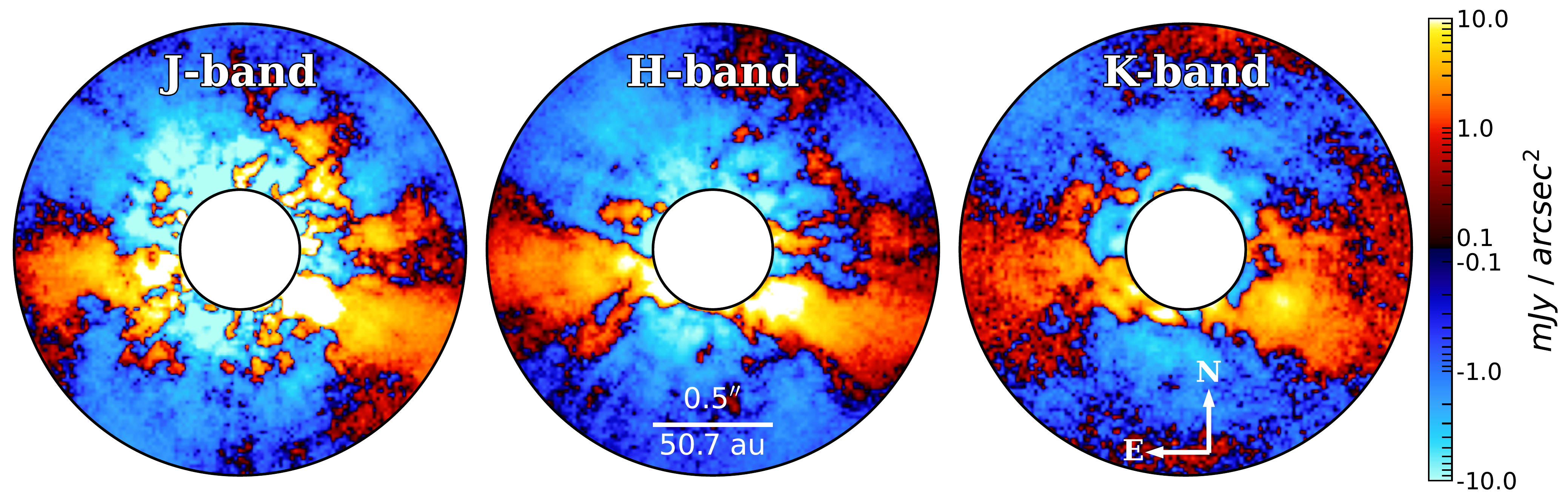}
    \caption{ADI-KLIP reduction of SCExAO/CHARIS HD 36546 data, with wavelength channels combined to produce images comparable to $J$ (channels $1-5$, $1.16-1.33$ $\micron$), $H$ (channels $8-14$, $1.47-1.80$ $\micron$) and $K$ (channels $16-21$, $1.93-2.29$ $\micron$) bands. Images are displayed with a symmetric logarithmic color scale which is linear in the range [-0.1, 0.1], and logarithmic otherwise.
    \label{fig:log_jhk}
    }
    \end{figure*}

    The SNRE of the disk is calculated following the standard procedure (e.g. \citealt{Goebel2018}); the signal per resolution element is determined by replacing each pixel with the sum of pixels within an aperture of one PSF full-width at half maximum (FWHM), while the noise per resolution element is the radial noise profile of the aforementioned signal with a finite-element-correction applied \citep{Mawet2014}. The quotient of these is taken to be the SNRE. The detection in the broadband images (Figure \ref{fig:snre}) is strongest for the ADI-ALOCI reduction, with a SNRE along the trace of the disk of $\sim 4-9$ for regions exterior to 0\farcs{}25\footnote{In the SNR calculation, we use a software mask to reduce the amount of disk signal included in the noise estimation.}. The ADI-KLIP reduction yields slightly weaker SNRE (typically $\sim 1$ less than the same locations in the ADI-ALOCI result). SNRE maps for the RDI-KLIP reduction suggest a somewhat weaker detection having SNRE of $\sim 3-4$ at small separations, dropping to $\sim 2$ beyond $0\farcs4$. We note, however, that the noise calculation for the RDI-KLIP reduction is more heavily affected by the disk's signal since RDI subtraction tends to be more conservative of disk flux.  Therefore, the resulting SNRE depends on the size of the noise mask utilized.  Using a slightly larger mask results in SNRE of $\sim 5-6$ out to $0\farcs6$, and $\sim 2.5$ beyond (Figure \ref{fig:snre}, upper right).
    The detection is clearest in H-band, where the disk appears much like the broadband result. Detections in J-band and K-band are weaker, with J-band suffering especially from prominent residual speckles at small separations, and K-band from weaker scattering and higher background flux (Figure \ref{fig:log_jhk}).

\section{Modeling the Debris Disk of HD 36546}
    \subsection{Disk Forward Modeling}\label{sec:forward_modeling}
    To probe the morphology of HD 36546's disk, we adopt a strategy of forward-modeling synthetic disk images as described in \citet{Currie2019, Lawson2020}, focused on KLIP instead of A-LOCI due to the former's much faster computational speed\footnote{In this case, KLIP is faster because reductions are performed in full annuli, requiring far fewer matrix operations than A-LOCI which is performed in much smaller annular segments.}. We create synthetic scattered-light images using a version of the GRaTeR software \citep{Augereau1999}. This model image is then rotated to an array of parallactic angles matching those of the data, and convolved with the instrumental PSF for each of the 22 CHARIS wavelength channels. Here, the PSF is estimated empirically by median combining normalized cutouts of the satellite spots for each wavelength over the full data sequence.  Using forward-modeling \citep{Pueyo2016}, we simulate annealing of the disk model for each CHARIS data cube in our sequence.  In this procedure, over-subtraction (where the presence of disk signal results in an over-bright PSF model), direct self-subtraction (inclusion of the disk signal in basis vectors), and indirect self-subtraction (perturbation of basis vectors by the disk signal) terms are computed from the Karhunen-Lo\`{e}ve modes  used for the science data reduction and their perturbations by the disk.   
    
    Once this process is completed, we compare the disk model to the wavelength-collapsed science image to assess the goodness of fit. For this purpose, we utilize $\chi^2_\nu$, which is calculated here within a region of interest as described in \citet{Lawson2020}. The region of interest considered is a rectangular box centered on the star with un-binned dimensions of 120 pixels $\times$ 50 pixels ($\sim 2\farcs0 \times 0\farcs8 $) and rotated 11\degr clockwise to approximately align the region with the disk's major axis. Models which are acceptably consistent with the overall best model's fitness,  $\chi^2_{\nu,min}$, are defined as those having $\chi^2_{\nu} \leq \chi^2_{\nu,min} + \sqrt{2 / \nu}$ \citep{Thalmann2013}.  Here, $\nu$ is the degree of freedom, defined as the difference between the number of bins in the region of interest when binned to resolution and the number of free model parameters. The disk models are restricted to those with simple one-ring geometries with linear flaring ($\beta = 1$), and a Gaussian vertical density distribution ($\gamma = 2$). 
    
    It is common to describe the angular distribution of scattered light in debris disks using the physically-unmotivated Henyey-Greenstein (H-G) scattering phase function \citep{Henyey1941}. The H-G SPF, parameterized by a single variable – the asymmetry parameter $g$\footnote{The asymmetry parameter is defined as the average of the cosine of the scattering angle, weighted by the (normalised) phase function, over all directions.}, can reproduce scattered light images at large scattering angles but often fails at smaller angles \citep[e.g.][]{Hughes2018}. Moreover, best-fit single H-G scattering asymmetry parameters are well correlated with the scattering angles probed by observations, while lacking correlation with the properties of the systems \citep{Hughes2018}. This appears to suggest that some other SPF, capable of reproducing observations irrespective of the scattering angles probed, is necessary. Linear combinations of two or more H-G phase functions of differing asymmetry parameter could provide a suitable solution specific to the details of a particular disk (e.g., its composition). However, optimization of a disk model with an SPF formed from the linear combination of $N$ H-G phase functions will introduce either $2N$ additional free parameters, with an asymmetry parameter ($g_i$) and weight ($w_i$) for each, or $2N-1$ free parameters for a constrained optimization problem (with the constraint that the weights sum to unity, allowing the weight for one asymmetry parameter to be assumed). Without providing additional constraints (e.g. that $w_i \geq w_{i+1}$), these weights and asymmetry parameters will also have significant correlations or degeneracies with one another. Since the utilized SPF is expected to alter the best-fit value of other disk parameters, the SPF cannot be optimized independently. Thus, exploration of such phase functions would introduce significant, possibly intractable, model complexity for attempts to optimize disk models.
    
     Another possibility, evidenced by measurements of both Solar-system and extrasolar dust, is that there exists a nearly universal SPF to describe circumstellar dust in a given wavelength regime \citep{Hughes2018}. In such a scenario, this SPF could be measured independently, such as by observation of Solar-system dust, and then simply adopted for use by groups modeling scattered light from disks. \citet{Lawson2020} showed that imagery of HD 15115 could be reasonably reproduced by adopting the fixed \citet{Hong1985} SPF -- modeled as the linear combination of three separate H-G phase functions based on observations of solar system zodiacal dust. Here, we again adopt the \citet{Hong1985} scattering phase function, with asymmetry parameters and weights fixed to the values identified in \citet{Hong1985}:  $g_1 = 0.7$, $g_2 = -0.2$, and $g_3 = -0.81$, with weights $w_1 = 0.665$, $w_2 = 0.330$, and $w_3 = 0.005$. 
    
    Altogether, our prescription results in a model having 6 free parameters:
    \begin{enumerate}
    \item $R_0$, the radius of peak grain density in au
    \item $\frac{H_0}{R_0}$, the ratio of disk scale height at $R_0$ to $R_0$
    \item $\alpha_{in}$, the power law index describing the change in radial density interior to $R_0$
    \item $\alpha_{out}$, the power law index describing the change in radial density exterior to $R_0$
    \item \textit{i}, the inclination of the disk in degrees
    \item \textit{PA}, the position angle of the disk in degrees
    \end{enumerate}
    
    To explore these parameters, we utilize the differential evolution (DE) optimization algorithm as described in \citet{Lawson2020}. In DE, a population of trial solutions is iteratively improved by adding the weighted difference of two random solutions to a third and keeping only those ``mutated'' population members that result in an improvement in the fitness metric. Algorithm meta-parameters are set as follows: $N_{pop} = 15$, resulting in a population size of 90 ($6$ free parameters $\times 15$), a dithered mutation constant $m= [0.5, 1.0]$, and a crossover probability $P = 0.7$. We test for convergence in the current population at the end of each generation. We consider the population to be converged if either of the following criteria is met:
    
    \begin{enumerate}
        \item For each varying parameter, the standard deviation among the set of normalized parameter values for the current population falls below a threshold value, $\rm \delta_{\sigma_X}$
        \item The standard deviation of the fitness of the current population, divided by the median fitness of the current population, falls below a threshold value, $\rm \delta_{\sigma_{FX}}$, and more than 10 generations have completed.
    \end{enumerate}
    
    Effectively, the first criterion is met by a population whose parameters have converged to a single minimum position, while the second accommodates a population with diverse parameter values but comparable goodness of fit, as might occur with weak or degenerate parameters. The additional requirement that 10 generations have finished for the second criteria prevents the unlikely case in which convergence might otherwise be achieved by a population in the first few generations with approximately uniform but weak fitness. Both thresholds, $\rm \delta_{\sigma_X}$ and $\rm \delta_{\sigma_{FX}}$, are set to 0.005\footnote{Typically, larger values $\sim 0.01$ are chosen. However, we opt for a stricter convergence requirement here purely for the sake of producing full corner plot visualizations -- the solutions show no statistically significant changes between a threshold of 0.01 and 0.005.}.
    
    We carry out several distinct procedures for optimizing the disk model. Our primary results focus on a procedure in which the model is optimized for both the RDI-KLIP and ADI-KLIP reductions simultaneously (i.e. seeking the model that minimizes the combined $\chi^2$ for both reductions). However, since the adopted noise profile for the RDI reduction suggests substantially larger noise levels than that of the ADI-KLIP reduction, its contribution to a combined $\chi^2_\nu$ metric will be minor; effectively, if the global best-fit solution differs considerably between the RDI and ADI reductions, this would not likely be apparent for such a procedure. As such, we also carry out procedures optimizing the model for each individual reduction. A secondary motivation for this is to validate the consistency of the solutions provided by DE. Toward this latter point, we also run two separate optimization procedures with identical settings for the RDI-KLIP data\footnote{Since DE involves a randomly initialized set of trial solutions which are then randomly mutated over subsequent generations, it is conceivable that two DE procedures could converge to distinct solutions.}.
    
    \subsection{Model Results}\label{sec:model_results}
\begin{figure*}%combined_cornerplot.pdf
    \includegraphics[width=\textwidth]{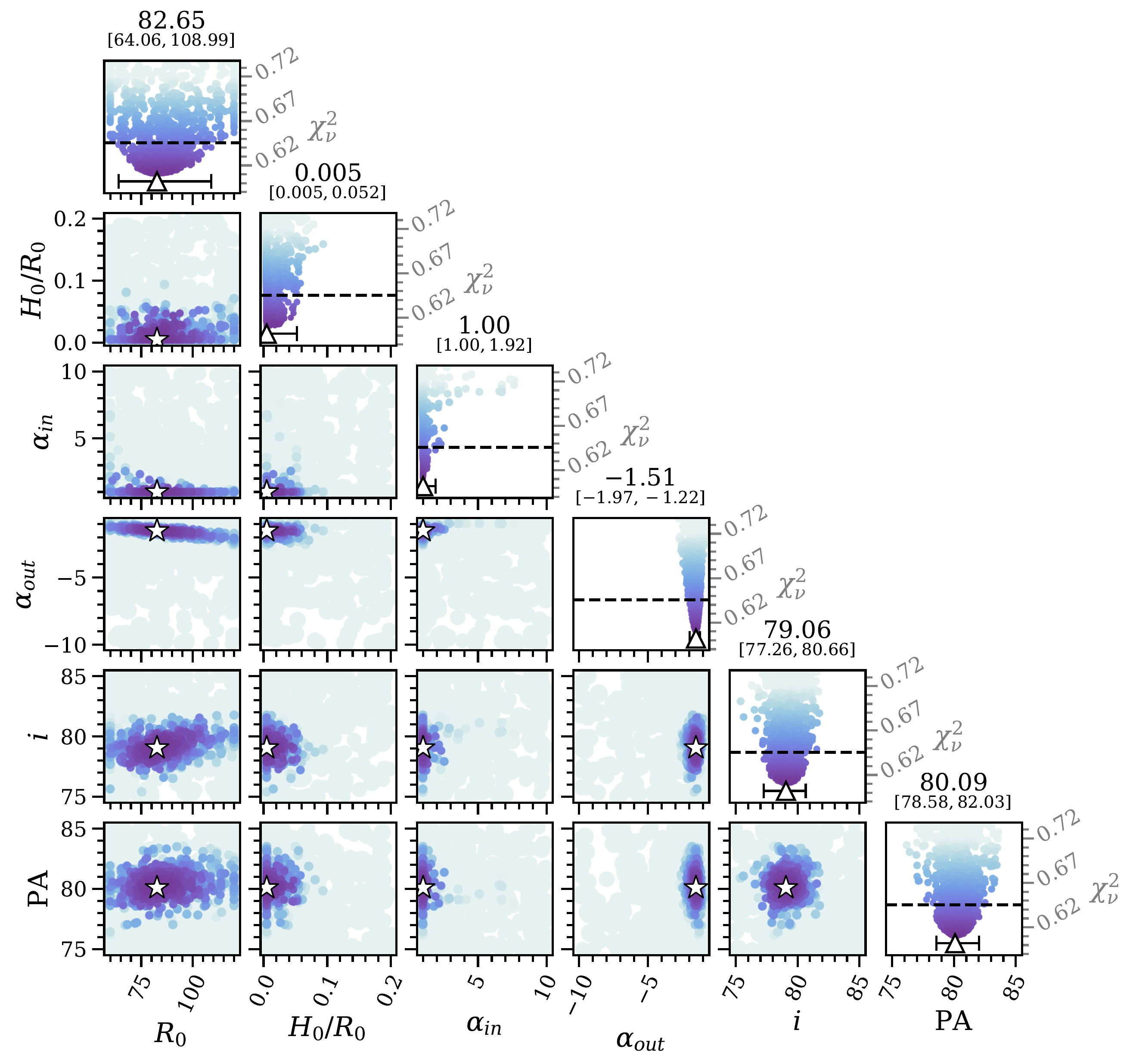}
    \caption{A variation of the corner plot for optimization of the disk model to both the RDI-KLIP and ADI-KLIP reductions simultaneously (corresponding to the row labeled `ADI \& RDI' in Table \ref{tab:model_results}). Each off-diagonal panel visualizes solutions as a function of two of the parameters. Each point in the subplots corresponds to a single sample, colored according to its $\chi^2_\nu$ score. The color mapping scales from $\chi^2_{\nu, min}$ to $\chi^2_{\nu, min} + 3\cdot \sqrt{2 / \nu}$ with darker colors indicating more optimal values of $\chi^2_\nu$. Diagonal elements provide a one-dimensional view of each of the parameters, while also serving as a color bar for the off-diagonals. The threshold fitness for acceptable solutions ($\chi^2_{\nu} \leq \chi^2_{\nu,min} + \sqrt{2 / \nu}$) is indicated by a horizontal black dashed line in each diagonal plot. The black-outlined star (triangle) marker in each off-diagonal (diagonal) panel indicates the position of the overall best-fitting model. The text at the top of each diagonal subplot provides the best-fitting value for the corresponding parameter, with the range beneath indicating the smallest and largest value producing an acceptable $\chi^2_{\nu}$; these limits are also depicted using error-bars for the `best-fit' marker in each diagonal panel.
    \label{fig:combined_corner}
    }
    \end{figure*}

\begin{figure*}%rdiklip_combrun_modelresults.pdf
    \includegraphics[width=\textwidth]{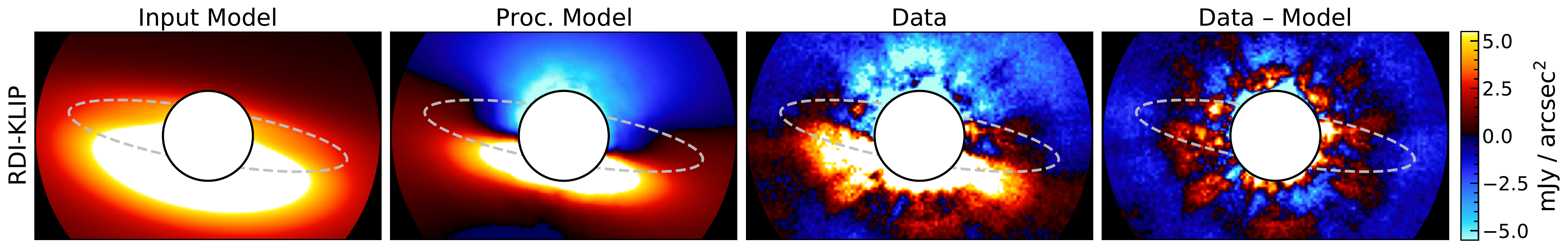}
    \caption{From left to right, for the RDI-KLIP reduction: the PSF-convolved input model, attenuated model, data, and residual for the overall best model resulting from the combined DE run. The grey dashed line overlaid in each panel is the ellipse corresponding to the radius, PA, and inclination of the model.
    \label{fig:rdiklip_model}
    }
    \end{figure*}

    The key results of the four modeling procedures (two for RDI-KLIP, one for ADI-KLIP, and one combined) are provided in Table \ref{tab:model_results}. A corner plot for the combined procedure is provided in Figure \ref{fig:combined_corner}. For analysis hereafter (e.g. for attenuation correction in Section \ref{sec:sb}), we adopt the best-fitting model resulting for the combined procedure, having parameters: $\rm R_0 = 82.65$ au, $H_0/R_0 = 0.005$, $\rm \alpha_{in} = 1.00$, $\rm \alpha_{out} = 1.51$, $\rm i = 79\fdg06$ and $\rm PA = 80\fdg09$ (visualized in Figures \ref{fig:rdiklip_model}, \ref{fig:adiklip_model}). The results of the other modeling procedures are visualized and compared in Appendix \ref{app:models}.
        
    \begin{deluxetable*}{cccccccc}
    \tablecaption{Disk Model Optimization Results}
    \tablehead{\colhead{} & \colhead{$R_{0}$ (au)} & \colhead{$H_{0} / R_{0}$} & \colhead{$\alpha_{in}$} & \colhead{$\alpha_{out}$} & \colhead{$i$ (deg)} & \colhead{\textit{PA} (deg)}  & \colhead{$\chi^2_{\nu}$}}
    \startdata
    \vspace{-0.2cm}  RDI-KLIP (1) & 75.4 & 0.007 & 1.0 & -1.4 & 78.9 & 79.0 & 0.45 \\
    $_{N=1980}$ & $_{[60.0,\;103.2]}$ & $_{[0.005,\;0.143]}$ & $_{[1.0,\;5.0]}$ & $_{[-2.1,\;-1.0]}$ & $_{[75.0,\;81.4]}$ & $_{[75.1,\;82.4]}$ & $_{< 0.50}$ \\
    \vspace{-0.35cm} & & & & & & & \\
    \tableline
    \vspace{-0.2cm}  RDI-KLIP (2) & 76.4 & 0.012 & 1.0 & -1.4 & 78.9 & 78.9 & 0.45 \\
    $_{N=2340}$ & $_{[60.0,\;102.7]}$ & $_{[0.005,\;0.138]}$ & $_{[1.0,\;5.4]}$ & $_{[-2.0,\;-1.0]}$ & $_{[75.0,\;81.9]}$ & $_{[75.5,\;82.4]}$ & $_{< 0.50}$ \\
    \vspace{-0.35cm} & & & & & & & \\
    \tableline
    \vspace{-0.2cm}  ADI-KLIP & 87.4 & 0.005 & 1.0 & -1.6 & 79.3 & 80.6 & 0.76 \\
    $_{N=1800}$ & $_{[70.3,\;114.6]}$ & $_{[0.005,\;0.050]}$ & $_{[1.0,\;1.9]}$ & $_{[-2.0,\;-1.3]}$ & $_{[76.9,\;81.3]}$ & $_{[78.7,\;82.6]}$ & $_{< 0.81}$ \\
    \vspace{-0.35cm} & & & & & & & \\
    \tableline
    \vspace{-0.2cm}  ADI \& RDI & 82.7 & 0.005 & 1.0 & -1.5 & 79.1 & 80.1 & 0.61 \\
    $_{N=2160}$ & $_{[64.1,\;109.0]}$ & $_{[0.005,\;0.052]}$ & $_{[1.0,\;1.9]}$ & $_{[-2.0,\;-1.2]}$ & $_{[77.3,\;80.7]}$ & $_{[78.6,\;82.0]}$ & $_{< 0.65}$ \\
    \vspace{-0.35cm} & & & & & & & \\
    \tableline
    \vspace{-0.2cm}  Currie et al. (2017) $^{a}$ & 75.6 & 0.118 & 3 & -3 & 75 & 75 & 1.03 \\
    $_{}$ & $_{[66.7,\;84.5]}$ & $_{[0.070,\;0.160]}$ & $_{[3,\;10]}$ & $_{[-3,\;-3]}$ & $_{[70,\;75]}$ & $_{}$ & $_{< 1.07}$ \\
    \vspace{-0.35cm} & & & & & & & \\
    \tableline
    Bounds & $[60,\;120]$ & $[0.005,\;0.2]$ & $[1,\;10]$ & $[-10,\;-1]$ & $[75,\;85]$ & $[75,\;85]$ & --\\
    \enddata
    \tablecomments{Results for each of the four distinct DE optimization runs outlined in Section \ref{sec:forward_modeling}, as well as the forward modeling of HD 36546 in \citet{Currie2017}. Each parameter column provides the value for the best model for the run indicated in the leftmost column (with $N$ indicating the total number of models evaluated for each procedure), with the following line providing the range of values resulting in an acceptably fitting model. The final column gives the corresponding run's minimum $\chi^2_\nu$ and the threshold for acceptable models. The permitted boundary for each parameter in modeling for CHARIS data (the modeling procedure of \citet{Currie2017} used different bounds) is given on the final line. We adopt the best model for the combined run (``ADI \& RDI'') for analysis hereafter. $a)$ In \citet{Currie2017}, models are discussed in terms of $\rm H_0$ ($\xi_0$ or $ksi_0$ in the text) rather than $\rm H_{0} / R_{0}$; for simplicity we present these results in terms of $\rm H_{0} / R_{0}$ instead. Additionally, we have updated relevant values for the difference in the assumed distance to the system — 114 pc in \citet{Currie2017}, versus 101.35 pc here. \label{tab:model_results}}
    \end{deluxetable*}
    
    \begin{figure*}%{adiklip_combrun_modelresults.pdf
    \includegraphics[width=\textwidth]{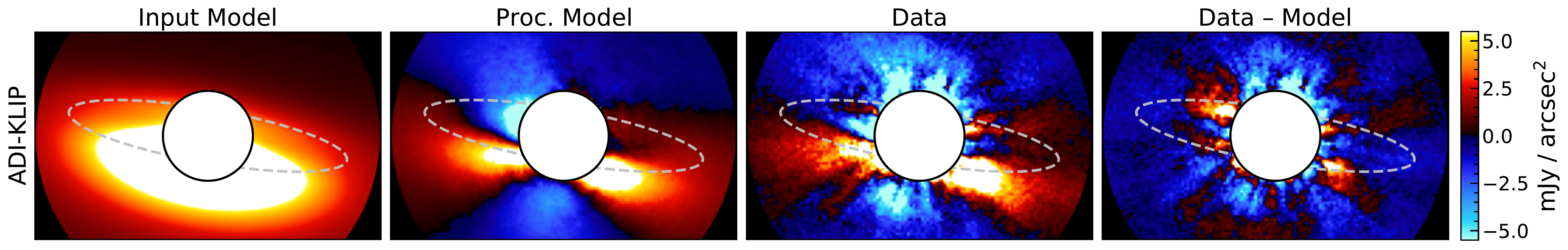}
    \caption{As Figure \ref{fig:rdiklip_model}, but for the ADI-KLIP reduction.
    \label{fig:adiklip_model}
    }
    \end{figure*}

    \subsection{Modeling Discussion}\label{sec:disc_modeling}
    The results of our forward-modeling procedure echo many of the themes of the results in \citet{Currie2017}: favoring a highly inclined disk model with shallow radial density profiles (small $\alpha_{in}$, $\alpha_{out}$), suggestive of an extended disk as opposed to the ring-like geometry often found for debris disks. 
    
    We emphasize here that, while the nominal parameter values are expected to very closely correspond to the global minimum of the parameter space, the acceptable parameter ranges from DE provide only a rough approximation of parameter uncertainties. Unlike ensemble samplers (e.g. Markov chain Monte Carlo), which draw samples from a parameter space in proportion to the statistical likelihood of those samples, DE is designed to rapidly approach the global minimum with little regard for sub-optimal combinations of parameters; in cases where the algorithm identifies a strong value for a particular parameter early in the run, the values of that parameter which yield ``acceptable'' results may be poorly explored. If a stronger approximation of parameter uncertainties is desired, repeated DE runs (or follow-up runs near the best DE solution with ``local'' optimization algorithms, such as damped least-squares) can provide thorough exploration of acceptable values while still requiring orders of magnitude fewer samples than would be required by an ensemble sampler.
    
    While highly inclined disks are expected to be relatively insensitive to changes in the $\alpha$ parameters (see, e.g., \citealt{Lawson2020}), these parameters are perhaps the best constrained of any in our modeling (see Figure \ref{fig:combined_corner}). This might be explained by the particularly small scale height favored by our modeling: $\rm H_0/R_0 = 0.005$, an order of magnitude smaller than the median value of 0.06 provided by \citet{Hughes2018} and the lower boundary for values permitted by our procedure. However, we note that scale heights up to $H_0/R_0 = 0.052$ are also acceptable solutions. In fact, the diagonal subplot for $H_0/R_0$ in Figure \ref{fig:combined_corner} shows that the minimum $\chi^2_\nu$ changes very little as a function of $H_0/R_0$ for $H_0/R_0 \lesssim 0.05$. Given that the corner plot shows no obvious strong correlations between $H_0/R_0$ and other parameters, this appears to suggest that our data simply cannot discriminate between $H_0/R_0$ values in this range. This is not particularly surprising, given that $H_0/R_0 = 0.05$ for a fiducial radius of $R_0 \sim 83$ au corresponds to $H_0 \sim 4.2$ au, which projects to $\sim 0\farcs04$ or $2.8$ pixels, compared to our resolution of $2.67$ pixels. Considering this apparent degeneracy for $H_0/R_0$, our results for this parameter should be interpreted as $H_0/R_0 \lesssim 0.05$. Additionally, we note that the value of $\alpha_{in}$ converges to the lower bound for permitted values ($\alpha_{in} = 1$). This result, indicating an exceptionally slow decline in density interior to $r_0$, could suggest that the disk is better described by a geometry with a nearly flat density slope over some intermediate region, followed by typical power-law slopes interior and exterior to that region. An alternative explanation for the strong $\alpha_{out}$ constraint that manifests here is addressed briefly in Section \ref{sec:spine}.
    
    The best-fit parameter values and acceptable ranges resulting from forward modeling in \citet{Currie2017} are included for comparison in Table \ref{tab:model_results}. Notably, this modeling was conducted on a coarse grid; the acceptable ranges provided are much less likely to correspond to complete ranges of acceptably-fitting parameters. Additionally, we note that, while the value for PA identified in \citet{Currie2017} is at the lower bound of allowed values for our presented DE runs, in earlier modeling runs not presented here the range of allowed PA values extended as low as 60$\degr$ but showed very poor fits for PA $\lesssim 75\degr$. As such, the final procedure restricted PA to the provided range to speed convergence. As stated previously, our results are thematically consistent with the picture of the disk suggested by the analysis in \citet{Currie2017}. Many of the specific parameter values identified, however, appear inconsistent; e.g., despite evaluating models with $\rm \alpha_{out}=-1$, the best-fit in \citet{Currie2017} is achieved for $\rm \alpha_{out}=-3$. Forward modeling the best fit solution from \citet{Currie2017} for our reductions results in a combined $\chi^2_\nu$ value of $0.85$ (well beyond the limit for acceptable models of $\chi^2_\nu = 0.65$). This apparent difference in identified optimal parameters might be explained by some combination of a few key differences in the modeling procedures: 
    \begin{enumerate}
        \item the data itself, e.g. with the CHARIS reduction making a partial detection of the fainter side of the disk where HiCIAO data did not,
        \item the scattering phase functions, with our choice to adopt the Hong phase function, as opposed to fitting an asymmetry parameter as in \citet{Currie2017}; in prior work modeling the disk of HD 15115 \citep{Lawson2020}, this was found to have a significant impact on the optimal parameters,
        \item the choice in \citet{Currie2017} to not vary PA, instead adopting the PA resulting from spine tracing,
        \item the parameter bounds, e.g. with values of $\alpha_{in}$ in \citet{Currie2017} not including the best-fit value that we identify,
        \item the parameter optimization, with \citet{Currie2017} identifying best-fit parameters through a much less thorough grid search.
    \end{enumerate}

    The best-fit model for the combined (``ADI \& RDI'') DE run results in a reduced $\chi^2_\nu$ score for each reduction that is very similar to the best scores achieved in the DE runs for individual reductions. The DE run for only the ADI-KLIP reduction results in a best-fit model having $\chi^2_\nu = 0.758$ (with ``acceptable'' models having $\chi^2_\nu \leq 0.808$), while the best-fit model of the combined run results in $\chi^2_\nu = 0.760$ when considering only the ADI-KLIP data — well within the limit to be considered an acceptable solution. Similarly, the DE runs for only the RDI-KLIP reduction result in a best-fit model having $\chi^2_\nu = 0.435$ (with $\chi^2_\nu \leq 0.504$ being acceptable) while the best-fit model of the combined run results in $\chi^2_\nu = 0.459$ for the RDI-KLIP reduction. In other words, the adopted disk model from our combined modeling run is also a reasonable explanation when considering the imagery of each data reduction individually.

\section{Disk Spine Trace and Surface Brightness}\label{sec:sb}
    To measure the disk's spine and surface brightness (SB), we proceed largely as detailed in \citet{Lawson2020} and summarized hereafter. In place of the RDI-KLIP reduction utilized for modeling, we perform spine fitting and surface brightness measurements on an RDI-KLIP reduction in which radial profile subtraction is utilized (with all other parameters remaining unchanged). This results in a generally weaker detection of the disk at very small separations and in wavelength collapsed imagery, but provides a more consistent detection across the field of view and in individual wavelength channels. For the sake of attenuation correction, we forward-model the adopted best-fitting model of Section \ref{sec:forward_modeling} for this additional reduction. The wavelength-collapsed image for this reduction and the corresponding forward modeling results are provided in Appendix \ref{app:rdi_rsub}.
    
    \subsection{Disk Spine}\label{sec:spine}
    The positions of peak radial brightness along the disk are referred to as the disk's spine, and are typically used as locations for surface brightness measurements and to approximate the geometric paramaters of the disk (e.g., fiducial radius, inclination, and position angle). We fit the spine position using wavelength-collapsed products and applying an initial image rotation of $-11\degr$ to approximately horizontally align the disk's major-axis. We then measure the spine at each pixel column position by fitting Lorentzian profiles to the image values weighted by the inverse of the radial noise profile. The fit Lorentzian peak and associated fit uncertainty are taken as the spine position and uncertainty. As a departure from the method of \citet{Lawson2020}, the fitting is carried out on both the ADI-KLIP and RDI-KLIP images simultaneously to identify the spine position most consistent with both images (rather than taking the weighted average of the results).
    
\begin{figure*}
    \includegraphics[width=\textwidth]{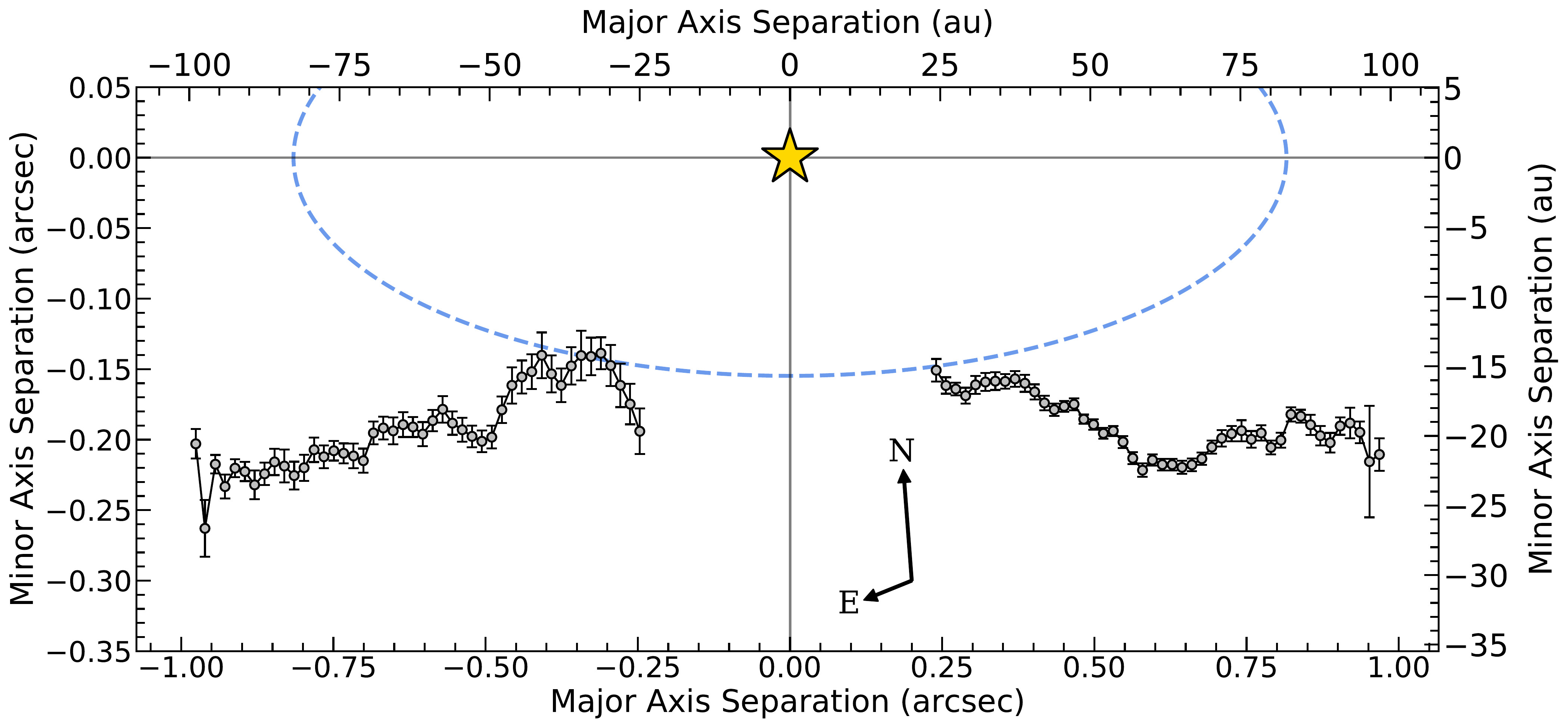}
    \caption{Stellocentric separation of disk spine fits along the major and minor axes (for a major axis position angle of 80\fdg09). The compass provided indicates both the orientation of the figure and the difference in scale between the axes (which results in $\textit{\textbf{N}} \not\perp \textit{\textbf{E}}$, as depicted).  The spine positions are identified by fitting Lorentzian profiles to both wavelength-collapsed KLIP images (ADI and RDI) simultaneously, and may indicate a swept-back wing profile (see Section \ref{sec:spine}). The dashed blue ellipse corresponds to the adopted best-fitting model of Section \ref{sec:forward_modeling} ($r_0 = 82.65$ au, $i = 79\fdg06$, \textit{PA}$ = 80\fdg09$), and is shown to demonstrate that the ellipsoidal profile very poorly describes the spine profile measured.
    \label{fig:spine}
    }
    \end{figure*}

    Notably, the fit spine (Figure \ref{fig:spine}) appears to grow further from the major axis at larger separations, suggesting a possible ``swept-back wing'' geometry (see e.g. \citealt{Hughes2018}), in which a disk's outer dust ring appears to bend (rather than remaining planar) at increasing separations. As a result of this, the measured spine is extremely poorly described by an ellipse and thus cannot easily provide a meaningful assessment of the disk's inclination or radius. While it may be possible to estimate the position angle of the disk's major axis from this spine profile, it is unclear how the value would be affected by the noted phenomenon. As such, we provide no estimate of the disk geometry based on spine fitting. For the locations of surface brightness measurements, we use the fit spine profile in place of the best-fit ellipse profile used for our analysis in \citet{Lawson2020} — correcting the initial rotation of $-11\degr$ using a rotational transformation of the spine positions.
    
    Interestingly, applying this spine fitting procedure to our best-fitting models also results in a divergent spine trace (both before and after attenuation is induced by forward modeling). After comparison to the complete sample of models, this behavior manifests only if $|\alpha_{out}| < 2$. This could suggest that the observed spine shape is simply the result of a disk with a dust distribution meeting this criteria and viewed nearly edge-on. Alternatively, models having such values of $\alpha_{out}$ may have been the only accessible means by which to reproduce this feature of the observations -- with the true cause being more similar to the explanations summarized by \citet{Hughes2018} (e.g. ISM interaction).

    \begin{figure*}
    \includegraphics[width=\textwidth]{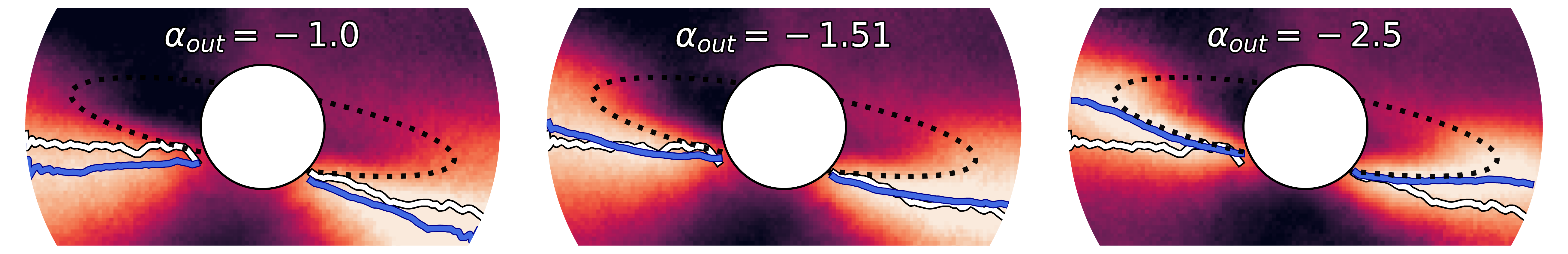}
    \caption{Disk models forward-modeled for our ADI-KLIP reduction with the parameters of the overall best model identified in Section \ref{sec:forward_modeling} but with differing values of $\alpha_{out}$. To emphasize the apparent disk spine, a $1/r^2$ correction and arbitrary color scale have been applied to each image. The dotted black ellipse in each panel corresponds to the radius, PA, and inclination of the model, while the blue line shows the spine trace found for the model following the procedure used for the real data (Section \ref{sec:spine}). The spine trace for the real data is overlaid in white. For $\alpha_{out} \gtrsim -2$, the fit spine deviates considerably from the profile of the ellipse. The model having $\alpha_{out} = -1.51$ corresponds to the overall best model identified in Section \ref{sec:forward_modeling}.
    \label{fig:model_spines}
    }
    \end{figure*}
    
    \subsection{Disk Surface Brightness}\label{sec:sb_measurement}
    
\begin{figure*}%sb.pdf
    \includegraphics[width=\textwidth]{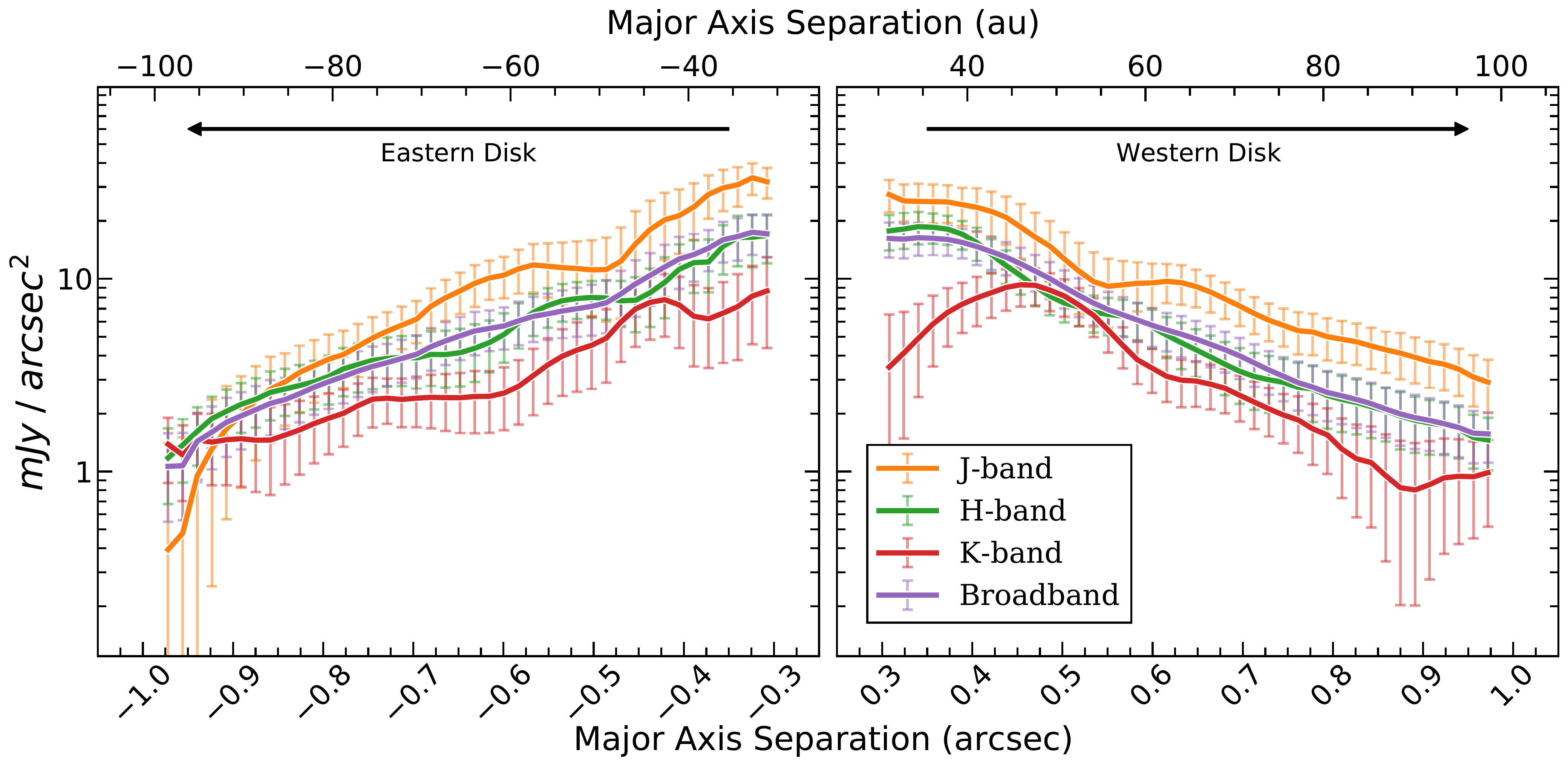}
    \caption{Measurements of peak surface brightness for HD 36546's disk as a function of projected stellocentric separation in arcseconds and au along the major axis in CHARIS data. Here, ``broadband'' refers to CHARIS's broadband. Since measurements are made using an aperture of radius 0\farcs06, comparable SB measurements can only be provided to within 0\farcs06 of the smallest separations at which the disk is recovered (~0\farcs25) — hence, SB is provided only to $\rho \sim 0\farcs31$. We note that the measurements here and in other Section \ref{sec:sb} figures are not statistically independent from neighboring measurements as a result of the 0\farcs12 diameter aperture. The apparent dip in western K-band surface brightness at small separations is not likely physical — given that no comparable behavior manifests in other bands, we conclude that this is probably the result of the somewhat compromised K-band data quality that was noted in Section \ref{sec:results}.
    \label{fig:sb}
    }
    \end{figure*}

    Peak surface brightness is measured for both the RDI-KLIP and ADI-KLIP reductions at the aforementioned spine locations using an aperture of diameter 0\farcs12 (7.4 pixels). As in \citet{Lawson2020}, a) we apply wavelength-dependent attenuation corrections based the adopted disk model (determined by comparison of the PSF-convolved input model cube and the attenuated model cube after forward modeling), and b) we approximate the uncertainty for each SB measurement as the standard deviation of an array of like-measurements taken at the same stellocentric radius but with azimuthal angles placed every 10\degr. These measurements are made on the ``broadband'' wavelength-collapsed images, as well as images corresponding to $J$, $H$, and $K$ bands. The final surface brightness profile for a given filter is taken to be the weighted average of the results for the two individual reductions. The results of this procedure are visualized in Figure \ref{fig:sb}.
    
\begin{figure*}%sb_surf_refl.pdf
    \includegraphics[width=\textwidth]{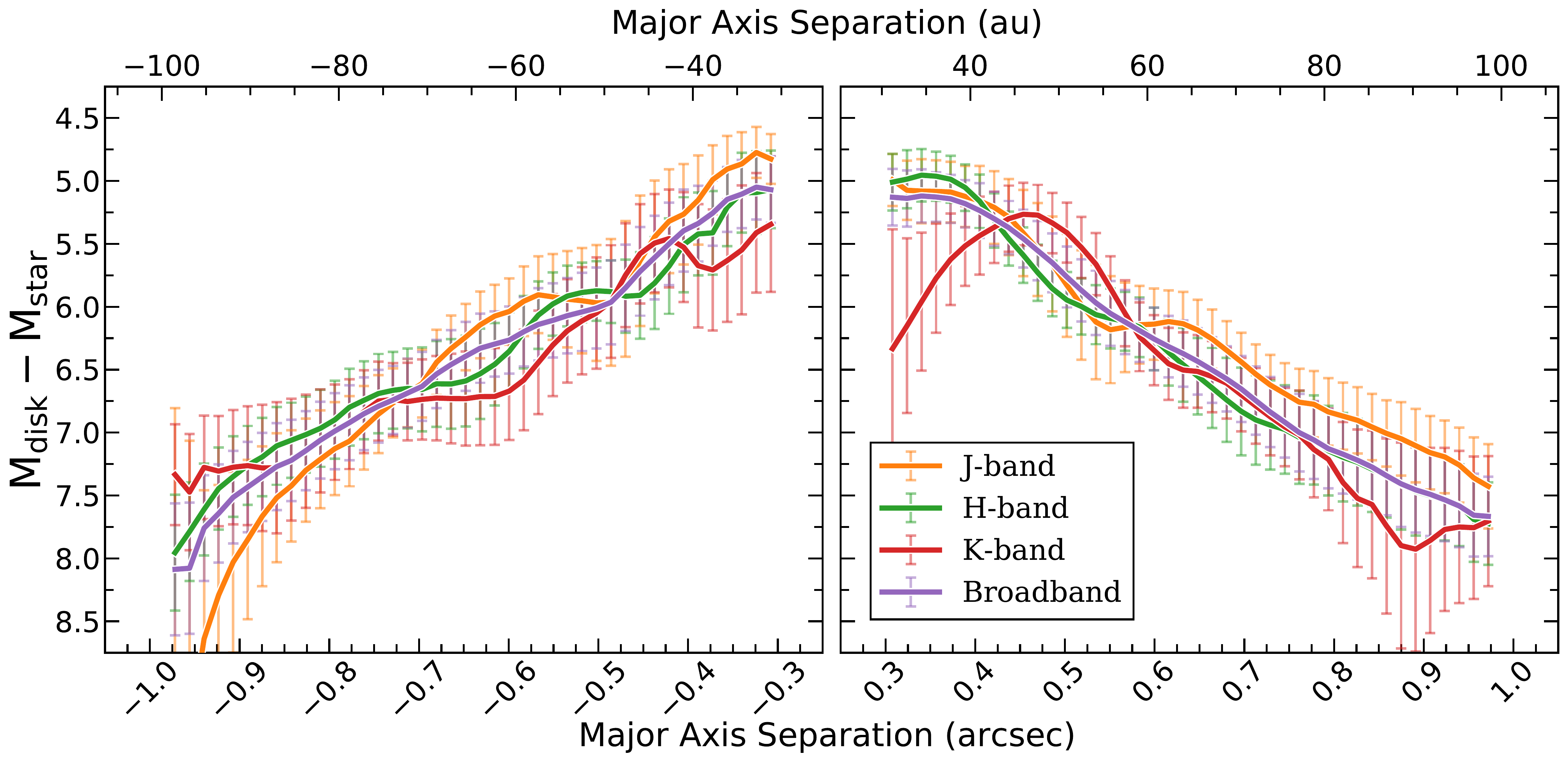}
    \caption{Measurements of surface contrast for HD 36546's disk in CHARIS data.
    \label{fig:sb_dmags}
    }
    \end{figure*}

    Disk surface contrast (the surface brightness of the disk relative to the stellar flux) is then computed by combining disk SB with measurements of the stellar flux for each filter (Figure \ref{fig:sb_dmags}).
    
\begin{figure}%disk_asymmetry.pdf
    \includegraphics[width=0.48\textwidth]{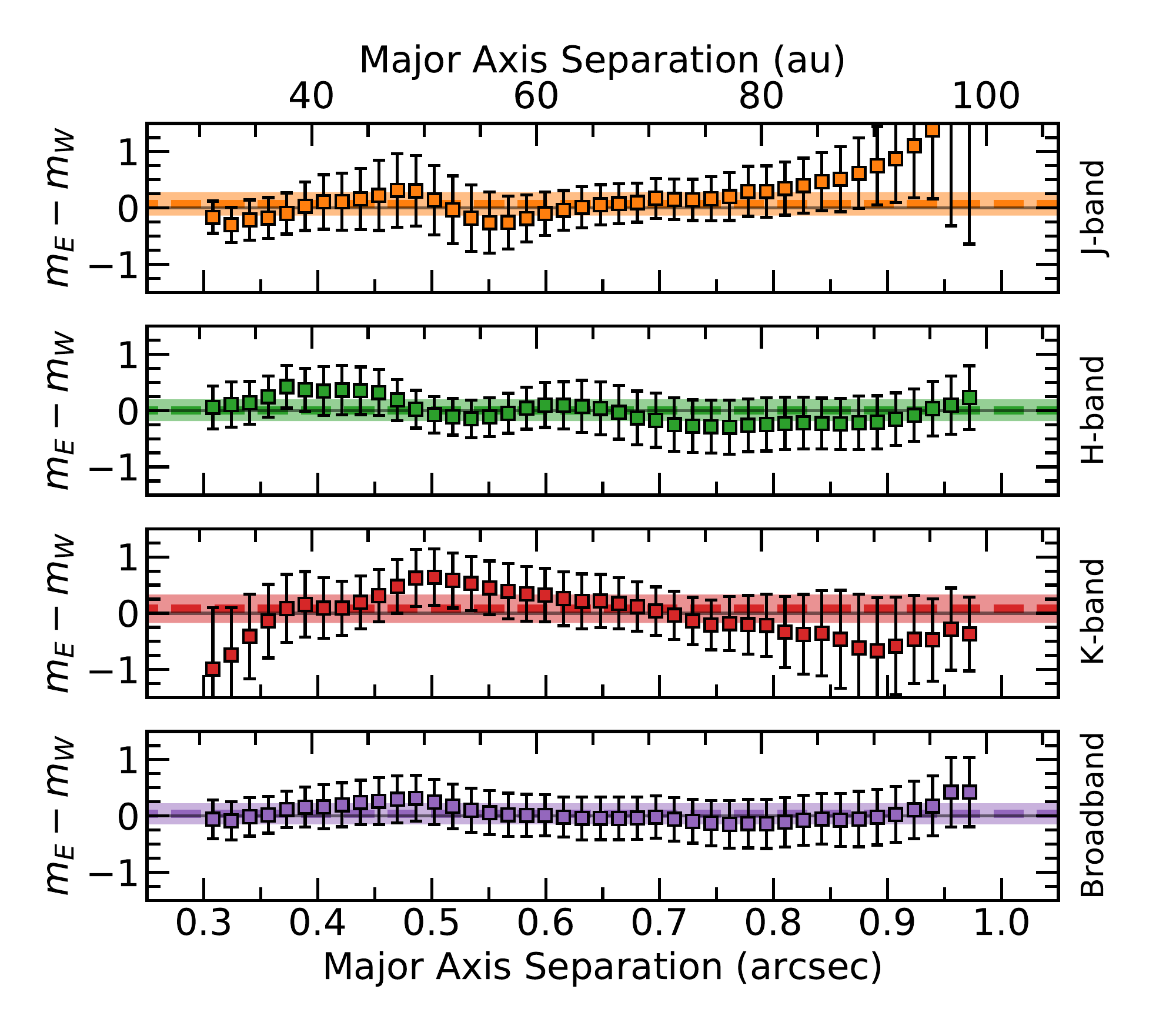}
    \caption{Relative surface brightness between the eastern and western extents of the disk for CHARIS $J$, $H$, $K$, and broadband. In each subplot, the horizontal colored dashed line indicates the weighted average of the full set of constituent flux measurements, while the colored region corresponds to the 3-$\sigma$ confidence interval for the average. The averages for all four filters are within 1-$\sigma$ of zero -- i.e. consistent with an overall symmetric disk flux.
    \label{fig:sb_asymmetry}
    }
    \end{figure}

    Any east-west asymmetry is measured by comparing the surface brightness of opposing sides in a given filter (Figure \ref{fig:sb_asymmetry}).
    
\begin{figure*}%jh_jk_color.pdf
    \includegraphics[width=\textwidth]{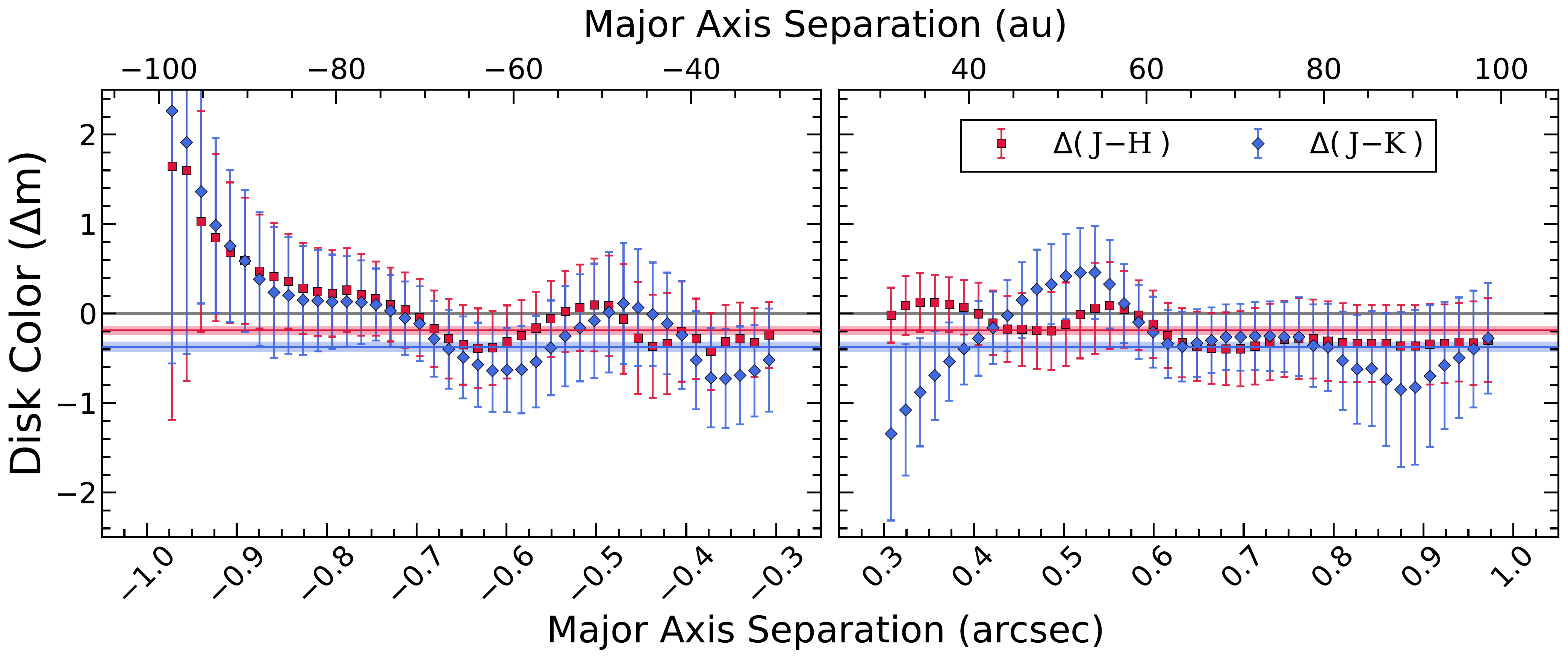}
    \caption{Disk color as a function of stellocentric separation along the disk major axis (see Section \ref{sec:sb}), with $\Delta(J-K)=(J-K)_{\rm{disk}}-(J-K)_{\rm{star}}$, etc. Major axis separation $< 0$ corresponds to the eastern side of the disk. Horizontal lines and shaded regions of the corresponding color represent the overall average and $1\sigma$ confidence respectively. The averages suggest an overall slightly blue color with $\geq 3\sigma$ confidence. $\Delta(H-K)$ manifests similarly to $\Delta(J-H)$.
    \label{fig:rel_color}
    }
    \end{figure*}

    Finally, the disk color in isolation from the stellar color is measured by converting surface contrasts to magnitudes and taking the difference of measurements between filters (see Figure \ref{fig:rel_color}), e.g. $(J_{\rm{disk}} - J_{\rm{star}}) - (K_{\rm{disk}} - K_{\rm{star}}) = (J-K)_{\rm{disk}}-(J-K)_{\rm{star}} = \Delta(J-K)$, where $\Delta(J-K)$ is the $J-K$ color of the disk without contribution from the star.

    \subsection{Surface Brightness Power-Law}\label{sec:sb_powerlaws}
\begin{figure}%sb_powerlaws.pdf
    \includegraphics[width=0.48\textwidth]{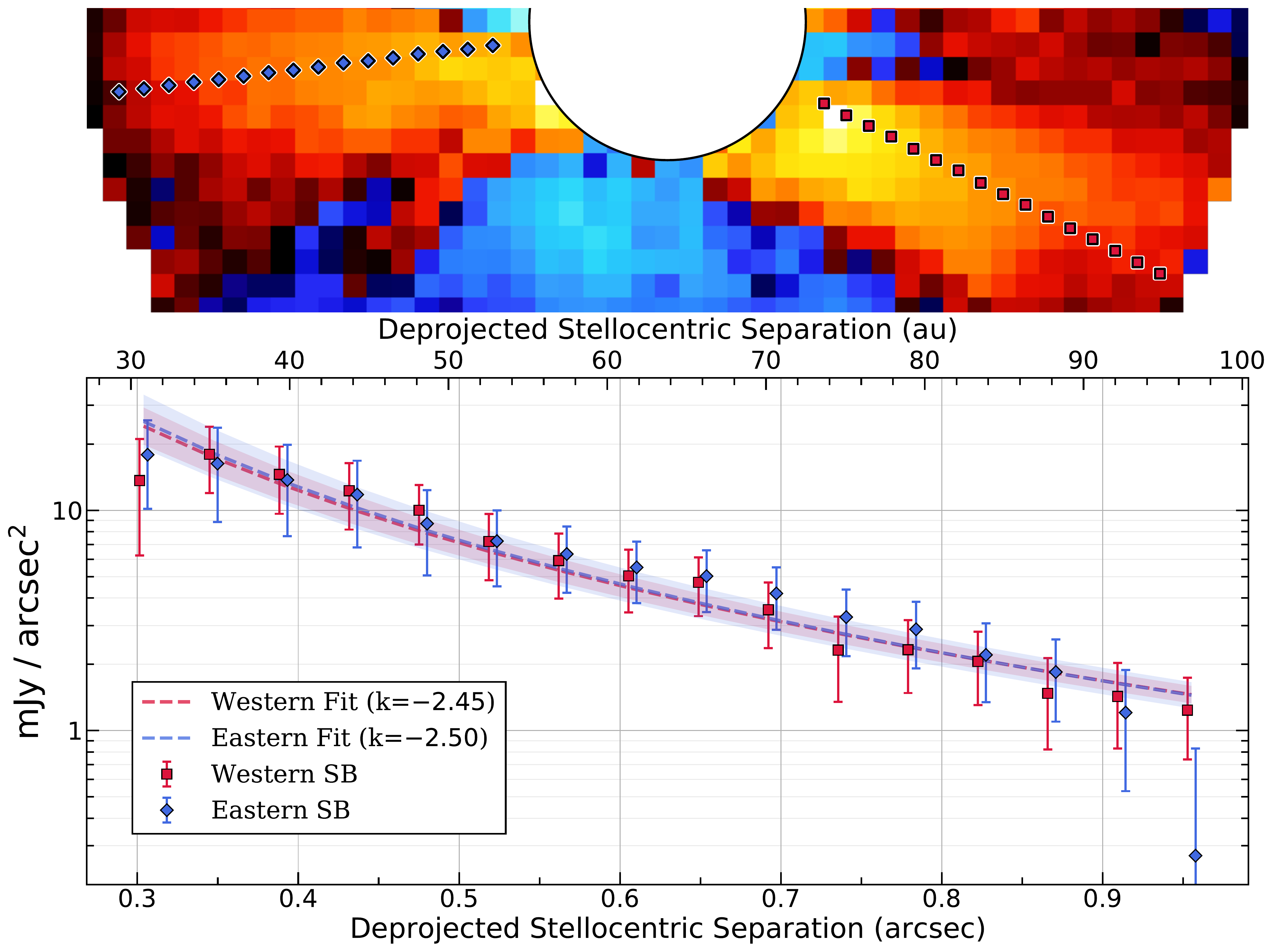}
    \caption{Broadband radial surface brightness (SB) profiles created following the procedure outlined in Section \ref{sec:sb_powerlaws}. The top panel shows the uncorrected ADI-KLIP image binned to resolution, with markers indicating the locations of SB measurements. The lower panel shows the radial surface brightness measurements, with east/west markers offset very slightly from one-another in the x-axis direction to avoid overlap. Power-law fits to the data are shown as dashed lines of the corresponding color (with shaded regions indicating the $1-\sigma$ confidence intervals). Separate sides of the disk are generally in good agreement, with confidence intervals overlapping significantly. \label{fig:sb_powerlaws}
    }
    \end{figure}

    For the broadband (wavelength-collapsed) images, we make an additional set of measurements to facilitate fitting a power-law to the disk's radial surface brightness profile. For this purpose, we follow the SB measurement procedure of Section  \ref{sec:sb_measurement}, except: 
    \begin{enumerate}
        \item measurements are made at positions with an array of radial separations but corresponding to a single scattering angle (25\degr\footnote{The particular scattering angle is chosen arbitrarily in an attempt to probe a scattering angle providing radial coverage with both generally high disk signal and well defined attenuation estimates in the east and west. Values within $\sim 5\degr$ provide comparable results.}, assuming $\textit{PA} = 80\fdg09$ and $\textit{i} = 79\fdg06$ based on the results of Section \ref{sec:model_results}), 
        \item images are binned to resolution ($\rm FWHM = 2.67$ pixels $=0\farcs043$),
        \item the aperture diameter is set to one resolution element
    \end{enumerate}
    The first change enables characterization of the radial dependence of the disk's SB for comparison with results in the literature, while the latter two changes result in approximately statistically independent SB measurements for fitting. The locations of these measurements are overlaid on the binned ADI-KLIP image in the top panel of Figure \ref{fig:sb_powerlaws}.  We then fit power-laws to these SB measurements as a function of deprojected stellocentric separation\footnote{Derived deprojected stellocentric separations consider only the determined PA and inclination of the system, and assume that our line of sight probes light scattered from the disk midplane. A more detailed assessment would take scale height, flaring, and similar effects into account, but is beyond the scope of this work.} for the eastern and western extents of the disk separately (Figure \ref{fig:sb_powerlaws}). We find best fitting power law slopes of $k=-2.45 \pm 0.14$ and $k=-2.50 \pm 0.21$ for the west and east respectively.

    \subsection{Surface Brightness Discussion}\label{sec:discussion_sb}
        \textbf{Disk Color} -- 
        Our individual measurements of disk color along the apparent spine of the disk (Figure \ref{fig:rel_color}) appear generally consistent with a disk scattering profile that is grey in NIR. However, taking the inverse-variance weighted average of the individual measurements for each color suggests a slightly blue color for the disk in both cases: $J-H = -0.19\pm0.05$ and $J-K = -0.37\pm0.06$. \footnote{This general result appears to be robust to the specific disk model adopted for the sake of attenuation correction. Calculating the average disk colors using each of the 691 acceptable disk models for attenuation correction instead resulted in color ranges of $J-H = [-0.23, -0.16]$ and $J-K = [-0.43, -0.32]$ with uncertainties comparable to the values found using the best-fit disk for attenuation correction.} Notably, the locations of the apparent disk spine at which this blue color was measured fall beyond the projected location of peak density for the disk. To assess the properties of the bulk population of dust in the disk, rather than just those of the dust pushed to larger separations by pressure from the parent star, we make additional measurements of the disk's color in the vicinity of the disk's peak radial density. For this purpose, we measure surface brightness in apertures of the same size (0\farcs06) along the brighter side of the disk at the projected location of the adopted disk model's density peak (i.e., along the southern edge of the ellipse overlaid in Figure \ref{fig:spine} -- corresponding to a single deprojected stellocentric separation). From these measurements, we proceed to average color measurements as before, finding: $J-H =  -0.13 \pm 0.08$, $H-K = -0.30 \pm 0.09$, and  $J-K = -0.40 \pm 0.11$. These results again suggest a slightly blue NIR color for the disk.
        
        Comparison with the numerical simulations in \citet{Boccaletti2003}, which model NIR disk colors as a function of the dust size distribution's minimum grain size ($\rm a_{min}$) and porosity (P), can enable some constraints regarding the size distribution of the dust in HD 36546's disk. With $\rm P=0$, for 1.1 $\micron$ versus 2.2 $\micron$ ($\sim J-K$), a blue disk color is achieved only for $\rm{a_{min}} \lesssim 0.25$ $\micron$. Similarly, for 1.6 $\micron$ versus 2.2 $\micron$ ($\sim H-K$) a blue color is achieved for $\rm{a_{min}} \lesssim 0.3$ \micron. Finally, for 1.1 $\micron$ versus 1.6 $\micron$ ($\sim J-H$), a blue color is achieved for $\rm{a_{min}} \lesssim 0.2$ \micron\footnote{The models of \citet{Boccaletti2003} are presented in terms of $\sim$ $J-K$ and $H-K$. For our purposes, we combine the provided measurements to present these results in terms of $J-H$ as well.} Based on these simulations, we can estimate a minimum grain size (assuming P=0) of $\rm a_{min} \lesssim 0.2$ $\micron$ to produce the blue color that we observe. The minimum grain size producing a blue color tends to increase for higher porosity, e.g. with $\rm a_{min} \lesssim 3$ $\micron$ producing a blue J$-$K color for P=0.95.
        
        \textbf{Disk Asymmetry} -- 
        Both individual measurements and the overall averages for east-west flux asymmetry show no statistically significant evidence for a disk that is asymmetric in brightness in any of the four filters analyzed (Figure \ref{fig:sb_asymmetry}). While the processed imagery of the disk may initially suggest some asymmetry is present -- with the western extent of the disk appearing brighter in all three presented reductions in Figure \ref{fig:log_reducs} -- no asymmetry is evident in SB measurements once attenuation corrections are applied. Thus, we conclude that the apparent asymmetry in CHARIS imagery is an artifact of PSF-subtraction and is not truly astrophysical in nature.
        
        \textbf{Disk Power-Law} -- 
        The results of power-law fitting (Figure \ref{fig:sb_powerlaws}) are in good agreement with the results of our asymmetry measurements (Figure \ref{fig:sb_asymmetry}) — again suggesting that the disk's flux is largely symmetrical at NIR wavelengths in the $\rho=0\farcs3–1\farcs0$ region of CHARIS's detection. 
        
        Comparing our best-fitting power law slope of $k \sim -2.5$ for HD 36546 ($\rho \sim 25 - 100$ au) to values in the literature for other debris disks provides further evidence that HD 36546 has a particularly shallow density slope. For the debris disk of HD 32297 ($\rho \sim 30 - 280$ au), \citet{Currie2012b} find $k \sim -6$ and for the debris disk of HR 4796 A ($\rho \sim 80 - 120$ au), \citet{Thalmann2011} report $k \sim -10$.

\section{Limits on Planets}\label{sec:planet_lims}

For the planet detection reduction outlined in Section \ref{sec:reduc}, we computed $5\sigma$ contrast limits in CHARIS broadband, following the procedure of \citet{Currie2018} for computing planet throughput corrections.  

Contrasts are about a factor of 2--3 worse at $\rho$ $<$ 0\farcs{}75 than good performance for bright diskless stars \citep{Currie2020spie} and slightly poorer than our previous limits for the debris disk-bearing HD 15115 \citep{Lawson2020}.   However, despite a loss of wavefront sensing sensitivity (see Section \ref{sec:data}), we reach contrasts at mid spatial frequencies slightly better than those achieved in \citet{Currie2017} with SCExAO/HiCIAO (e.g. $4.4\times10^{-6}$ at 0\farcs{}45 with CHARIS, versus $\sim8\times 10^{-6}$ with HiCIAO).   This gain is likely due to the use of SDI for additional speckle suppression.   

To convert from contrast to planet mass limits, we integrated the hot-start, solar metallicity, hybrid cloud, synthetic planet spectra provided by \citet{Spiegel2012} over the CHARIS bandpass and divided by the integrated stellar spectrum.  We consider two mappings between contrast and mass, corresponding to the lower and upper age limits estimated for the system (3 Myr and 10 Myr).  

\begin{figure}%contrast_plus_hgca.pdf
\includegraphics[width=0.48\textwidth]{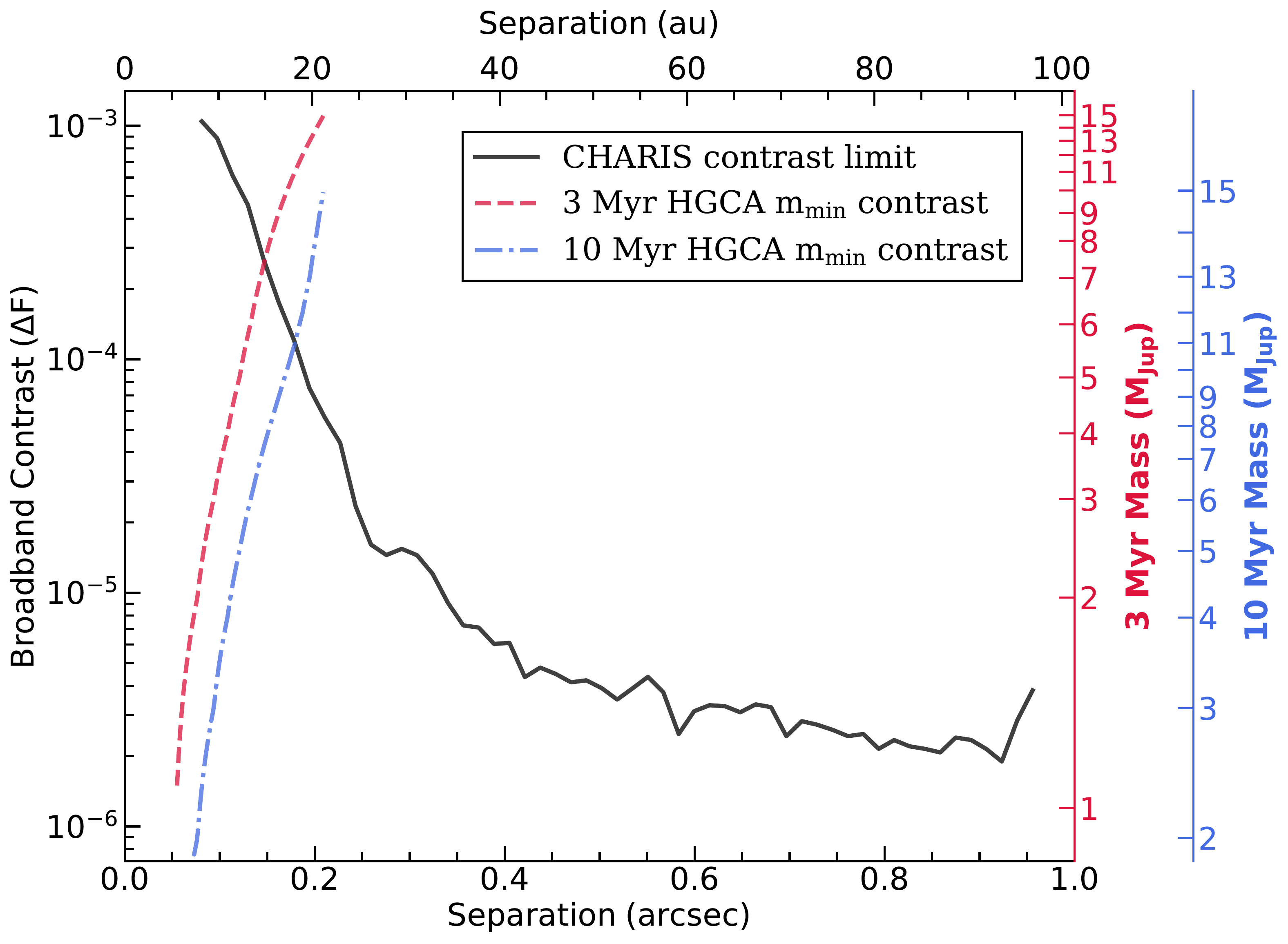}
\caption{Broadband ($1.13-2.39$ $\micron$) contrast curve (solid black line) for the planet detection reduction of CHARIS HD 36546 data, as outlined in Section \ref{sec:reduc}. $5 \sigma$ contrast is given as a function of stellocentric angular separation (arcsec, lower x-axis) and projected separation (au, upper x-axis). The right edge of the figure gives contrast mapped to companion masses for 3 Myr (red) and 10 Myr (blue) planets, based on hot-start, solar-metallicity, hybrid cloud planet evolution models of \citet{Spiegel2012}. Overplotted lines of the corresponding color indicate the minimum mass and projected separations corresponding to the companion implied by HGCA measurements, mapped to contrasts based on the same models. The displayed ages represent the lower and upper limits of age estimates for HD 36546.
\label{fig:contrast_curve}
}
\end{figure}

The right axis of Figure \ref{fig:contrast_curve} denotes these mass limits. For $r \gtrsim 35$ au ($\rho$ $\sim0\farcs35$), our reduction is sensitive to $\sim2$ $\rm M_{Jup}$ planets at $\sim3$ Myr, and $\sim5$ $\rm M_{Jup}$ planets at $\sim10$ Myr. Despite achieving generally inferior contrasts compared to the best reduction of CHARIS data for the debris disk-bearing star HD 15115 \citep{Lawson2020}, we are able to probe slightly less massive planets at a given projected separation due to HD 36546's relative youth.

While our data do not result in the detection of any candidate planets, HD 36546 may show tentative, indirect evidence for a massive and (in-principle) imageable companion. The \textit{Hipparcos-Gaia} Catalog of Accelerations (HGCA; \citealt{Brandt2018}) provides calibrated proper motion data for $\sim 115,000$ nearby stars spanning the $\sim 24$ year baseline between \textit{Hipparcos} and \textit{Gaia} astrometry. Accelerations based on this data have been used to measure dynamical masses of directly imaged exoplanets \citep[e.g.][]{Brandt2019}, as well as to provide evidence of gravitational interaction with yet-unseen companions for targeted direct imaging searches \citep{Currie2020b,Steiger2021}.   Brandt et al. (2021, submitted) updates HGCA, replacing \textit{Gaia} DR2 measurements with those from \textit{Gaia} eDR3, which yield a factor of $\sim$ 3 improvement in astrometric precision.   Using these updated \textit{Gaia} measurements, HGCA for HD 36546 (Table \ref{tab:hgca_measurements}) indicates a marginally significant acceleration relative to a model of constant proper motion — $\chi^2 = 7.65$, or 2.3$\sigma$ for two degrees of freedom — that \textit{could} be indicative of a companion.

    \begin{deluxetable}{@{\extracolsep{-2pt}}cccccccc}
    \tablewidth{0pt}
    \tablecaption{HGCA Measurements of HD 36546}
    \tablehead{
    \colhead{Source} & \colhead{$\mu_\alpha$} & \colhead{$\sigma_\alpha$} & \colhead{$\mu_\delta$} & \colhead{$\sigma_\delta$} & \colhead{Corr} & \colhead{$t_\alpha$} & \colhead{$t_\delta$}
    }
    \startdata
    \textit{Hip} & 8.513 & 0.974 & -41.876 & 0.472 & 0.257 & 1991.15 & 1990.76 \\
    \textit{HG} & 7.368  & 0.028   & -41.226 & 0.014  & 0.087 & &  \\
    \textit{Gaia} & 7.508  & 0.056 & -41.305 & 0.040 & -0.137  & 2016.00  & 2016.50 \\
    \enddata
    \tablecomments{HGCA calibrated proper motion measurements for HD 36546 \citep{Brandt2018} as updated with \textit{Gaia} eDR3 measurements in Brandt et al. (2021, submitted).
    }\label{tab:hgca_measurements}
    \end{deluxetable}
    
Following equations in \citet{Brandt2019} to estimate companion mass as a function of projected separation using HGCA accelerations, we can assess the completeness of our search for the HD 36546 companion that would induce this acceleration. Since radial velocity (RV) measurements are not available for HD 36546, we adopt an RV acceleration of zero to compute a minimum mass instead. Mapping of the resulting minimum masses to contrasts for 3 Myr and 10 Myr are overlaid on Figure \ref{fig:contrast_curve}. These results suggest that we would likely recover the HGCA companion for a projected separation beyond $\sim 0\farcs15$ assuming an age of 3 Myr, or $\sim 0\farcs18$ for an age of 10 Myr. 

\section{Conclusions and Future Work}\label{sec:conclusion}

Our study provides the first multi-wavelength imaging of the HD 36546 disk and recovers disk signal to separations as small as $\sim 0\farcs25$. Through detailed modeling implementing a \citet{Hong1985} scattering phase function and using the differential evolution optimization algorithm, we provide an updated schematic of HD 36546's disk. This schematic suggests a disk with a particularly shallow radial dust density profile, a fiducial radius of $\sim 83$ au, an inclination of $\sim 79\degr$, and a position angle of $\sim 80 \degr$. Through spine tracing, we find a disk spine that is consistent with our modeling results, but also consistent with a ``swept-back wing'' geometry. Surface brightness measurements show no significant flux asymmetry between the eastern and western extents of the disk, and slightly blue NIR colors on average. While we report no evidence of directly imaged companions, we provide contrast limits to constrain the most plausible remaining companions.

Notably, prior work modeling the debris disks of HD 36546 and HD 15115 in NIR total intensity data identified very distinct optimal HG SPF asymmetry parameters of $g = 0.70–0.85$ \citep{Currie2017} and $g = 0.3-0.4$ \citep{Engler2019} respectively. Alongside the results of \citet{Lawson2020}, we have shown that imagery of both disks can be reasonably reproduced by models using the same \citet{Hong1985} SPF instead. As is suggested in \citet{Hughes2018}, this may support the existence of a nearly universal SPF for circumstellar dust. Additional data to study smaller scattering angles and additional disk systems is necessary to clarify this possibility. If such an SPF is verified, it will enable both better exploration of model parameters (by reducing the number of varying parameters) and tighter constraints on parameter values (by mitigating parameter degeneracies).

Deeper follow-up observations with SCExAO/CHARIS would provide a) further constraints on companions within the disk and b) higher SNR measurements of the disk color, enabling a more detailed assessment of the dust properties of the disk.  The upcoming replacement of Subaru's AO188 system and additional commissioned improvements to wavefront sensing are expected to improve SCExAO/CHARIS contrasts at small separations by roughly a factor of 10 by late 2021 \citep{Currie2020spie}. Assuming a system age of 3 Myr, these contrasts would enable detection of $\sim$1 $\rm M_{Jup}$-mass planets at $\sim 25$ au.   A future non-detection would likely restrict the location of any super-Jovian mass planet to $\rho \lesssim 0\farcs1$. 

Observations measuring the polarized intensity of the disk at comparable scales -- or the total intensity of the disk in a wider field of view -- could clarify the underlying cause of the noted diverging spine trace. CHARIS's integral field spectropolarimetry mode or space-based imaging of the system could provide these data. Further, polarized intensity imaging would enable even tighter constraints on disk model parameters by better probing regions of the disk where the polarized intensity phase function results in more favorable sensitivity.

\acknowledgements

We thank our referee, whose comments helped us to improve both the content and clarity of this manuscript.

This research is based on data collected at Subaru Telescope, which is operated by the National Astronomical Observatory of Japan. We are honored and grateful for the opportunity of observing the Universe from Maunakea, which has cultural, historical and natural significance in Hawaii.

We wish to acknowledge the critical importance of the current and recent Subaru telescope operators, daycrew, computer support, and office staff employees.  Their expertise, ingenuity, and dedication is indispensable to the continued successful operation of Subaru.  

The development of SCExAO was supported by the Japan Society for the Promotion of Science (Grant-in-Aid for Research \#23340051, \#26220704, \#23103002, \#19H00703 \& \#19H00695), the Astrobiology Center of the National Institutes of Natural Sciences, Japan, the Mt Cuba Foundation and the director’s contingency fund at Subaru Telescope.  We acknowledge funding support from the NASA XRP program via grants 80NSSC20K0252 and NNX17AF88G.  T.C. was supported by a NASA Senior Postdoctoral Fellowship.  M.T. is supported by JSPS KAKENHI grant Nos.18H05442, 15H02063, and 22000005. K.W. acknowledges support from NASA through the NASA Hubble Fellowship grant HST-HF2-51472.001-A awarded by the Space Telescope Science Institute, which is operated by the Association of Universities for Research in Astronomy, Incorporated, under NASA contract NAS5-26555.

\appendix
\section{Additional Model Results}\label{app:models}
\begin{figure*}
    \includegraphics[width=\textwidth]{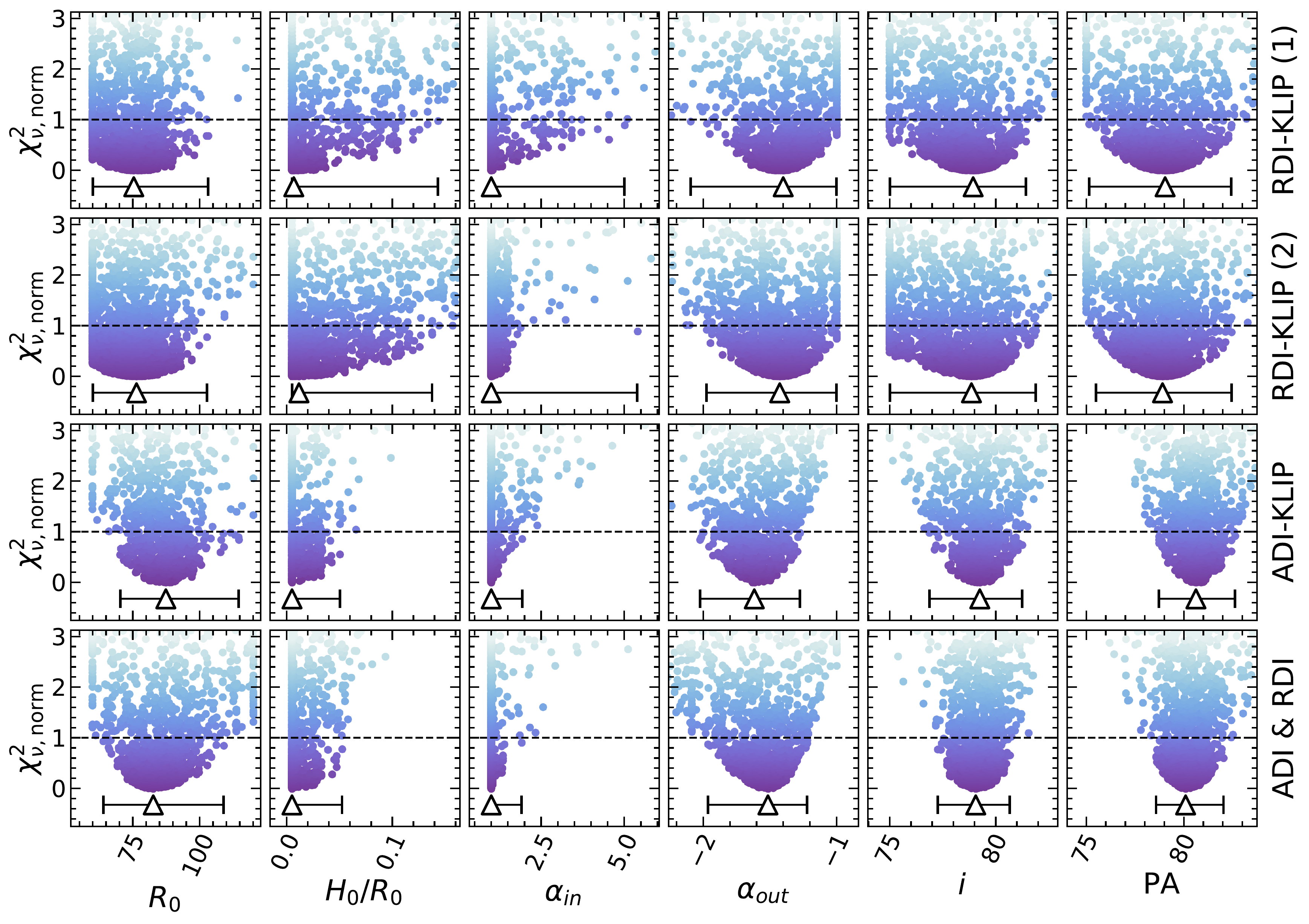}
    \caption{
    The diagonal (one dimensional) elements of corner plots for each of the four DE procedures (as indicated to the right of each row of subplots), as summarized in Table \ref{tab:model_results}. $\chi^2_\nu$ is presented normalized as $\chi^2_{\nu,norm} = (\chi^2_\nu - \chi^2_{\nu,min}) / \sqrt{2/\nu}$, such that the best model for each run falls at $\chi^2_{\nu,norm}=0$ and acceptable models have $\chi^2_{\nu,norm}\leq1$. Parameter bounds are truncated to better show the distributions in the vicinity of the minima.
    \label{fig:corner_diagonals}
    }
    \end{figure*}

The results of our additional differential evolution disk model optimization procedures are included here, with the information of Table \ref{tab:model_results} graphically summarized in Figure \ref{fig:corner_diagonals}. Individual reductions are in close agreement with one another; while the optimal values for some parameters (e.g. $R_0$) differ noticeably compared to the permitted parameter bounds, each reduction's best fit values fall within the acceptable ranges for the other reductions. 

Comparison of the samples for $\alpha_{in}$ between the two RDI-KLIP DE runs (the first two rows of Figure \ref{fig:corner_diagonals}) illustrates our prior warning regarding approximation of parameter uncertainties from DE (see Section \ref{sec:disc_modeling}). While the runs ultimately yield very similar acceptable ranges, the comparable upper limit achieved for ``RDI-KLIP (2)'' is the result of a single sample; except for the lone acceptable model with $\alpha_{in}\sim 5.5$, the acceptable range of $\alpha_{in}$ values for this run would manifest much more comparably to that of the ADI-KLIP run.

\begin{figure*}
    \includegraphics[width=\textwidth]{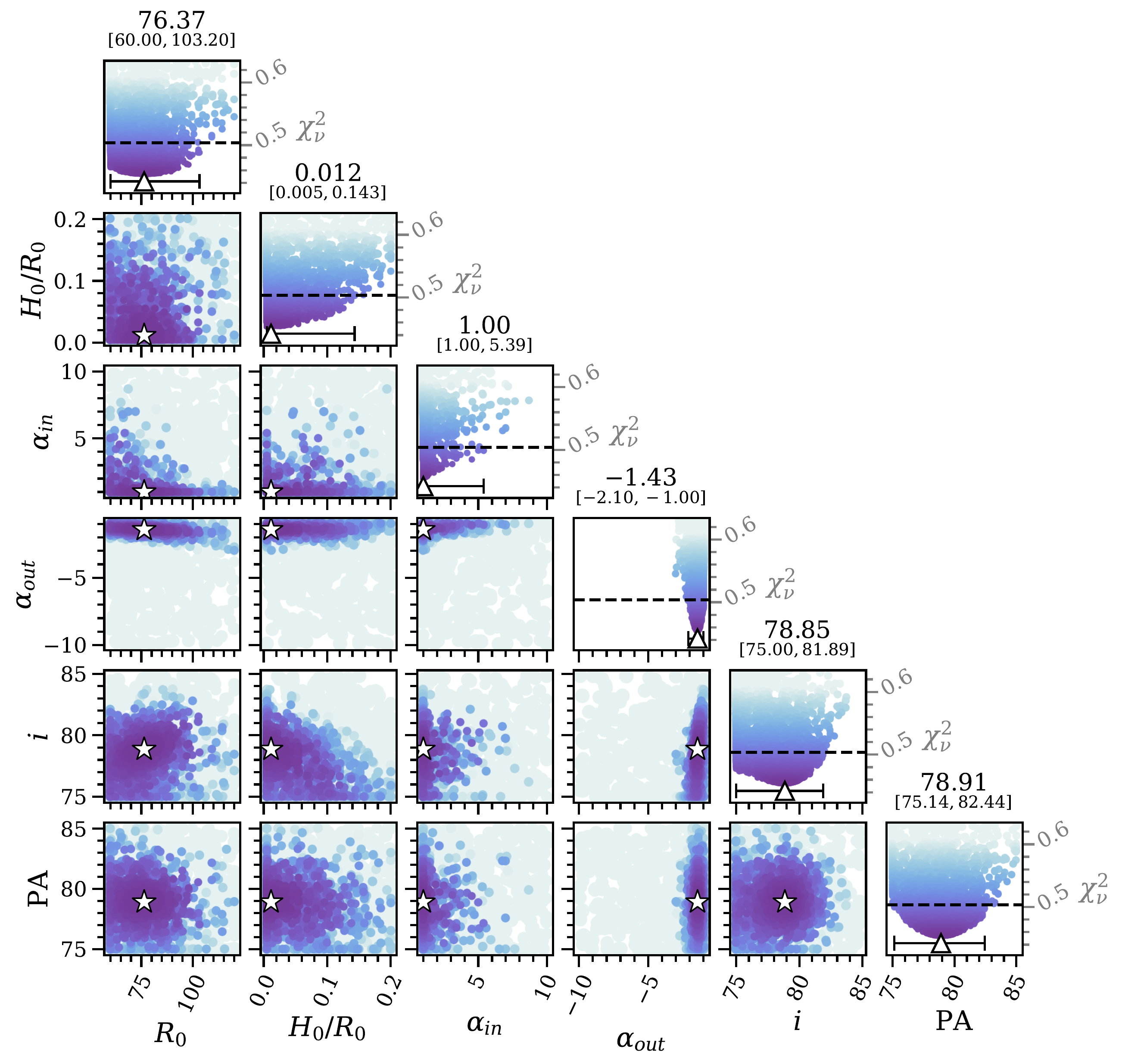}
    \caption{
    As Figure \ref{fig:combined_corner}, but for optimization of the disk model to only the RDI-KLIP reduction of HD 36546. This figure includes both runs for the RDI-KLIP data (corresponding to the entries for ``RDI-KLIP (1)'' and ``RDI-KLIP (2)'' in Table \ref{tab:model_results}).
    \label{fig:rdi_corner}
    }
    \end{figure*}

\begin{figure*}
    \includegraphics[width=\textwidth]{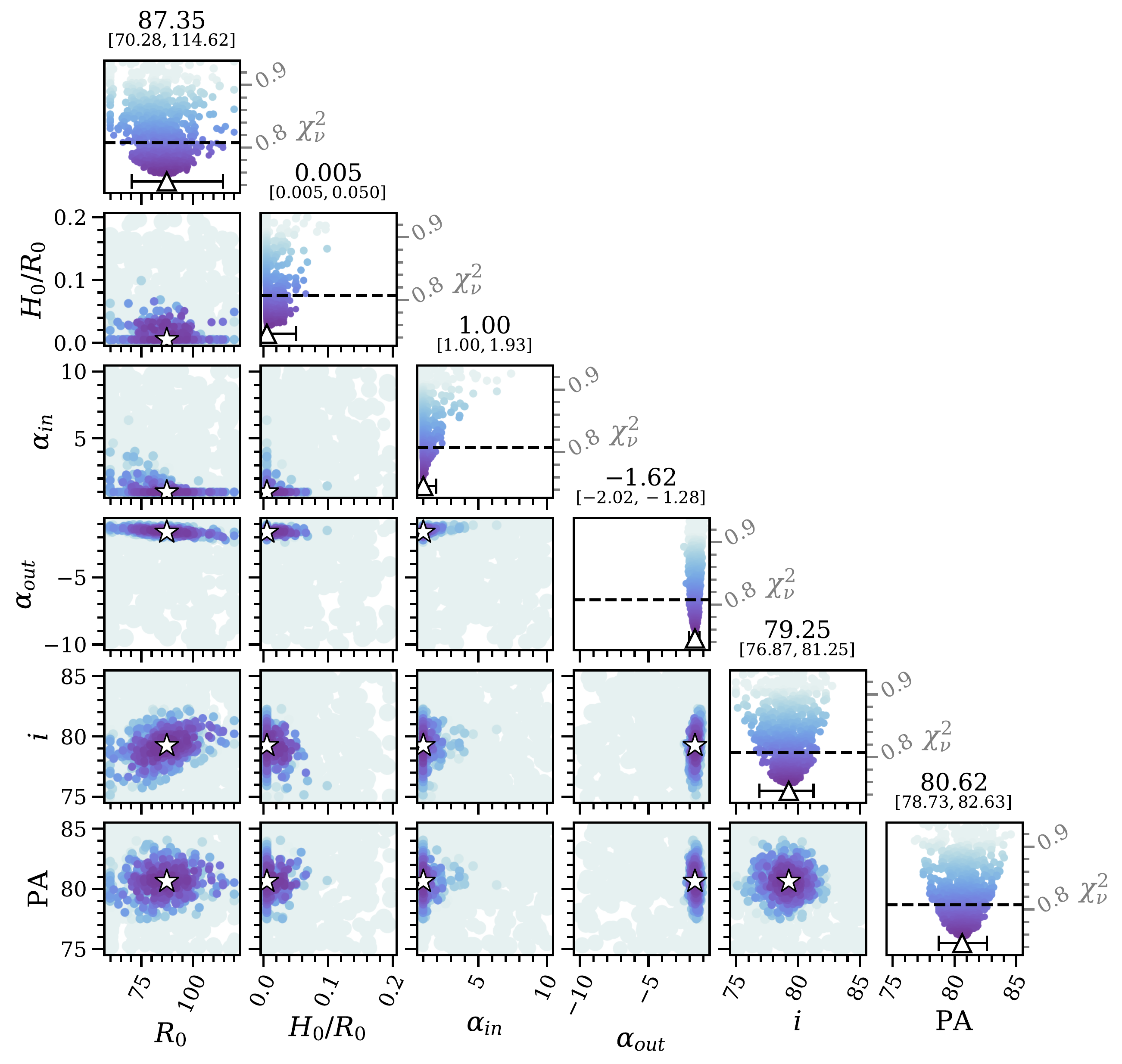}
    \caption{As Figure \ref{fig:combined_corner}, but for optimization of the disk model to only the ADI-KLIP reduction of HD 36546 (corresponding to the entry for ``ADI-KLIP'' in Table \ref{tab:model_results}). Compared to that of the RDI-KLIP data, the ADI-KLIP optimization results in stronger constraints for the model parameters by virtue of a higher signal-to-noise ratio.
    \label{fig:adiklip_corner}
    }
    \end{figure*}

As a result of the larger noise levels in the RDI-KLIP reduction, the RDI-KLIP modeling procedure (Fig. \ref{fig:rdi_corner}) shows much weaker constraints than the ADI-KLIP (Fig. \ref{fig:adiklip_corner}) modeling procedure. Despite some differences in the resulting parameter values, the best-fit solutions for these procedures manifest very similarly to those of the combined run (e.g. Figures \ref{fig:rdiklip_model}, \ref{fig:adiklip_model}).

\section{RDI-KLIP Reduction using Radial Profile Subtraction}\label{app:rdi_rsub}
\begin{figure*}
    \includegraphics[width=\textwidth]{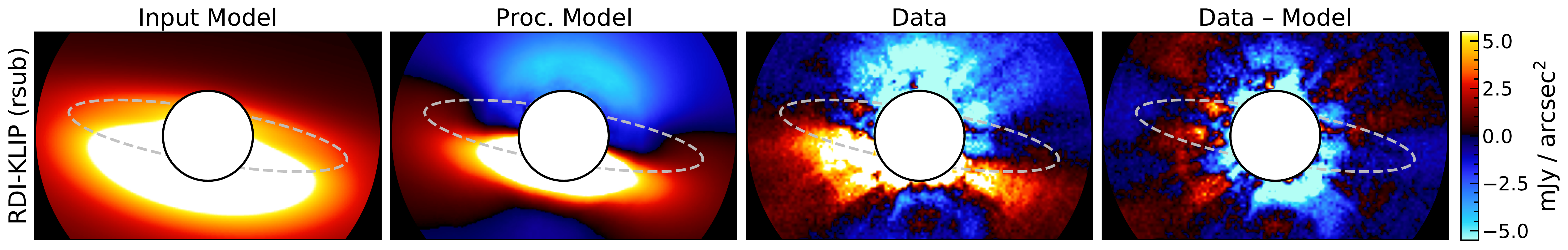}
    \caption{As Figure \ref{fig:rdiklip_model}, but for the RDI-KLIP reduction with radial profile subtraction that was utilized for spine tracing and surface brightness measurements (Section \ref{sec:sb}).
    \label{fig:rdikliprsub_model}
    }
    \end{figure*}

The results of forward modeling the adopted best-fitting model for the radial profile subtracted RDI-KLIP reduction (see Section \ref{sec:sb}) are provided in Figure \ref{fig:rdikliprsub_model}.

\bibliography{refs}{}

\begin{thebibliography}{}
\expandafter\ifx\csname natexlab\endcsname\relax\def\natexlab#1{#1}\fi
\providecommand{\url}[1]{\href{#1}{#1}}
\providecommand{\dodoi}[1]{doi:~\href{http://doi.org/#1}{\nolinkurl{#1}}}
\providecommand{\doeprint}[1]{\href{http://ascl.net/#1}{\nolinkurl{http://ascl.net/#1}}}
\providecommand{\doarXiv}[1]{\href{https://arxiv.org/abs/#1}{\nolinkurl{https://arxiv.org/abs/#1}}}

\bibitem[{{Augereau} {et~al.}(1999){Augereau}, {Lagrange}, {Mouillet},
  {Papaloizou}, \& {Grorod}}]{Augereau1999}
{Augereau}, J.~C., {Lagrange}, A.~M., {Mouillet}, D., {Papaloizou}, J.~C.~B.,
  \& {Grorod}, P.~A. 1999, \aap, 348, 557.
\newblock \doarXiv{astro-ph/9906429}

\bibitem[{{Beuzit} {et~al.}(2019){Beuzit}, {Vigan}, {Mouillet}, {Dohlen},
  {Gratton}, {Boccaletti}, {Sauvage}, {Schmid}, {Langlois}, {Petit},
  {Baruffolo}, {Feldt}, {Milli}, {Wahhaj}, {Abe}, {Anselmi}, {Antichi},
  {Barette}, {Baudrand}, {Baudoz}, {Bazzon}, {Bernardi}, {Blanchard}, {Brast},
  {Bruno}, {Buey}, {Carbillet}, {Carle}, {Cascone}, {Chapron}, {Charton},
  {Chauvin}, {Claudi}, {Costille}, {De Caprio}, {de Boer}, {Delboulb{\'e}},
  {Desidera}, {Dominik}, {Downing}, {Dupuis}, {Fabron}, {Fantinel}, {Farisato},
  {Feautrier}, {Fedrigo}, {Fusco}, {Gigan}, {Ginski}, {Girard}, {Giro},
  {Gisler}, {Gluck}, {Gry}, {Henning}, {Hubin}, {Hugot}, {Incorvaia}, {Jaquet},
  {Kasper}, {Lagadec}, {Lagrange}, {Le Coroller}, {Le Mignant}, {Le Ruyet},
  {Lessio}, {Lizon}, {Llored}, {Lundin}, {Madec}, {Magnard}, {Marteaud},
  {Martinez}, {Maurel}, {M{\'e}nard}, {Mesa}, {M{\"o}ller-Nilsson}, {Moulin},
  {Moutou}, {Orign{\'e}}, {Parisot}, {Pavlov}, {Perret}, {Pragt}, {Puget},
  {Rabou}, {Ramos}, {Reess}, {Rigal}, {Rochat}, {Roelfsema}, {Rousset}, {Roux},
  {Saisse}, {Salasnich}, {Santambrogio}, {Scuderi}, {Segransan}, {Sevin},
  {Siebenmorgen}, {Soenke}, {Stadler}, {Suarez}, {Tiph{\`e}ne}, {Turatto},
  {Udry}, {Vakili}, {Waters}, {Weber}, {Wildi}, {Zins}, \&
  {Zurlo}}]{Beuzit2019}
{Beuzit}, J.~L., {Vigan}, A., {Mouillet}, D., {et~al.} 2019, \aap, 631, A155,
  \dodoi{10.1051/0004-6361/201935251}

\bibitem[{{Boccaletti} {et~al.}(2003){Boccaletti}, {Augereau}, {Marchis}, \&
  {Hahn}}]{Boccaletti2003}
{Boccaletti}, A., {Augereau}, J.~C., {Marchis}, F., \& {Hahn}, J. 2003, \apj,
  585, 494, \dodoi{10.1086/346019}

\bibitem[{{Brandt}(2018)}]{Brandt2018}
{Brandt}, T.~D. 2018, \apjs, 239, 31, \dodoi{10.3847/1538-4365/aaec06}

\bibitem[{{Brandt} {et~al.}(2019){Brandt}, {Dupuy}, \& {Bowler}}]{Brandt2019}
{Brandt}, T.~D., {Dupuy}, T.~J., \& {Bowler}, B.~P. 2019, \aj, 158, 140,
  \dodoi{10.3847/1538-3881/ab04a8}

\bibitem[{{Brandt} {et~al.}(2017){Brandt}, {Rizzo}, {Groff}, {Chilcote},
  {Greco}, {Kasdin}, {Limbach}, {Galvin}, {Loomis}, {Knapp}, {McElwain},
  {Jovanovic}, {Currie}, {Mede}, {Tamura}, {Takato}, \& {Hayashi}}]{Brandt2017}
{Brandt}, T.~D., {Rizzo}, M., {Groff}, T., {et~al.} 2017, Journal of
  Astronomical Telescopes, Instruments, and Systems, 3, 048002,
  \dodoi{10.1117/1.JATIS.3.4.048002}

\bibitem[{{Chen} {et~al.}(2020){Chen}, {Mazoyer}, {Poteet}, {Ren},
  {Duch{\^e}ne}, {Hom}, {Arriaga}, {Millar-Blanchaer}, {Arnold}, {Bailey},
  {Bruzzone}, {Chilcote}, {Choquet}, {De Rosa}, {Draper}, {Esposito},
  {Fitzgerald}, {Follette}, {Hibon}, {Hines}, {Kalas}, {Marchis}, {Matthews},
  {Milli}, {Patience}, {Perrin}, {Pueyo}, {Rajan}, {Rantakyr{\"o}}, {Rodigas},
  {Roudier}, {Schneider}, {Soummer}, {Stark}, {Wang}, {Ward-Duong},
  {Weinberger}, {Wilner}, \& {Wolff}}]{Chen2020}
{Chen}, C., {Mazoyer}, J., {Poteet}, C.~A., {et~al.} 2020, \apj, 898, 55,
  \dodoi{10.3847/1538-4357/ab9aba}

\bibitem[{{Cloutier} {et~al.}(2014){Cloutier}, {Currie}, {Rieke}, {Kenyon},
  {Balog}, \& {Jayawardhana}}]{Cloutier2014}
{Cloutier}, R., {Currie}, T., {Rieke}, G.~H., {et~al.} 2014, \apj, 796, 127,
  \dodoi{10.1088/0004-637X/796/2/127}

\bibitem[{{Currie} {et~al.}(2015{\natexlab{a}}){Currie}, {Cloutier},
  {Brittain}, {Grady}, {Burrows}, {Muto}, {Kenyon}, \& {Kuchner}}]{Currie2015}
{Currie}, T., {Cloutier}, R., {Brittain}, S., {et~al.} 2015{\natexlab{a}},
  \apj, 814, L27, \dodoi{10.1088/2041-8205/814/2/L27}

\bibitem[{{Currie} {et~al.}(2008){Currie}, {Kenyon}, {Balog}, {Rieke}, {Bragg},
  \& {Bromley}}]{Currie2008}
{Currie}, T., {Kenyon}, S.~J., {Balog}, Z., {et~al.} 2008, \apj, 672, 558,
  \dodoi{10.1086/523698}

\bibitem[{{Currie} {et~al.}(2009){Currie}, {Lada}, {Plavchan}, {Robitaille},
  {Irwin}, \& {Kenyon}}]{Currie2009}
{Currie}, T., {Lada}, C.~J., {Plavchan}, P., {et~al.} 2009, \apj, 698, 1,
  \dodoi{10.1088/0004-637X/698/1/1}

\bibitem[{{Currie} {et~al.}(2015{\natexlab{b}}){Currie}, {Lisse}, {Kuchner},
  {Madhusudhan}, {Kenyon}, {Thalmann}, {Carson}, \& {Debes}}]{Currie2015b}
{Currie}, T., {Lisse}, C.~M., {Kuchner}, M., {et~al.} 2015{\natexlab{b}},
  \apjl, 807, L7, \dodoi{10.1088/2041-8205/807/1/L7}

\bibitem[{{Currie} {et~al.}(2011){Currie}, {Burrows}, {Itoh}, {Matsumura},
  {Fukagawa}, {Apai}, {Madhusudhan}, {Hinz}, {Rodigas}, {Kasper}, {Pyo}, \&
  {Ogino}}]{Currie2011}
{Currie}, T., {Burrows}, A., {Itoh}, Y., {et~al.} 2011, \apj, 729, 128,
  \dodoi{10.1088/0004-637X/729/2/128}

\bibitem[{{Currie} {et~al.}(2012{\natexlab{a}}){Currie}, {Debes}, {Rodigas},
  {Burrows}, {Itoh}, {Fukagawa}, {Kenyon}, {Kuchner}, \&
  {Matsumura}}]{Currie2012}
{Currie}, T., {Debes}, J., {Rodigas}, T.~J., {et~al.} 2012{\natexlab{a}},
  \apjl, 760, L32, \dodoi{10.1088/2041-8205/760/2/L32}

\bibitem[{{Currie} {et~al.}(2012{\natexlab{b}}){Currie}, {Rodigas}, {Debes},
  {Plavchan}, {Kuchner}, {Jang-Condell}, {Wilner}, {Andrews}, {Kraus}, {Dahm},
  \& {Robitaille}}]{Currie2012b}
{Currie}, T., {Rodigas}, T.~J., {Debes}, J., {et~al.} 2012{\natexlab{b}}, \apj,
  757, 28, \dodoi{10.1088/0004-637X/757/1/28}

\bibitem[{{Currie} {et~al.}(2017){Currie}, {Guyon}, {Tamura}, {Kudo},
  {Jovanovic}, {Lozi}, {Schlieder}, {Brandt}, {Kuhn}, {Serabyn}, {Janson},
  {Carson}, {Groff}, {Kasdin}, {McElwain}, {Singh}, {Uyama}, {Kuzuhara},
  {Akiyama}, {Grady}, {Hayashi}, {Knapp}, {Kwon}, {Oh}, {Wisniewski}, {Sitko},
  \& {Yang}}]{Currie2017}
{Currie}, T., {Guyon}, O., {Tamura}, M., {et~al.} 2017, \apj, 836, L15,
  \dodoi{10.3847/2041-8213/836/1/L15}

\bibitem[{{Currie} {et~al.}(2018){Currie}, {Brandt}, {Uyama}, {Nielsen},
  {Blunt}, {Guyon}, {Tamura}, {Marois}, {Mede}, {Kuzuhara}, {Groff},
  {Jovanovic}, {Kasdin}, {Lozi}, {Hodapp}, {Chilcote}, {Carson}, {Martinache},
  {Goebel}, {Grady}, {McElwain}, {Akiyama}, {Asensio-Torres}, {Hayashi},
  {Janson}, {Knapp}, {Kwon}, {Nishikawa}, {Oh}, {Schlieder}, {Serabyn},
  {Sitko}, \& {Skaf}}]{Currie2018}
{Currie}, T., {Brandt}, T.~D., {Uyama}, T., {et~al.} 2018, \aj, 156, 291,
  \dodoi{10.3847/1538-3881/aae9ea}

\bibitem[{{Currie} {et~al.}(2019){Currie}, {Marois}, {Cieza}, {Mulders},
  {Lawson}, {Caceres}, {Rodriguez-Ruiz}, {Wisniewski}, {Guyon}, {Brandt},
  {Kasdin}, {Groff}, {Lozi}, {Chilcote}, {Hodapp}, {Jovanovic}, {Martinache},
  {Skaf}, {Lyra}, {Tamura}, {Asensio-Torres}, {Dong}, {Grady}, {Gerard},
  {Fukagawa}, {Hand}, {Hayashi}, {Henning}, {Kudo}, {Kuzuhara}, {Kwon},
  {McElwain}, \& {Uyama}}]{Currie2019}
{Currie}, T., {Marois}, C., {Cieza}, L., {et~al.} 2019, \apjl, 877, L3,
  \dodoi{10.3847/2041-8213/ab1b42}

\bibitem[{{Currie} {et~al.}(2020{\natexlab{a}}){Currie}, {Guyon}, {Lozi},
  {Sahoo}, {Vievard}, {Deo}, {Chilcote}, {Groff}, {Brandt}, {Lawson}, {Skaf},
  {Martinache}, \& {Kasdin}}]{Currie2020spie}
{Currie}, T., {Guyon}, O., {Lozi}, J., {et~al.} 2020{\natexlab{a}}, in Society
  of Photo-Optical Instrumentation Engineers (SPIE) Conference Series, Vol.
  11448, Society of Photo-Optical Instrumentation Engineers (SPIE) Conference
  Series, 114487H, \dodoi{10.1117/12.2576349}

\bibitem[{{Currie} {et~al.}(2020{\natexlab{b}}){Currie}, {Brandt}, {Kuzuhara},
  {Chilcote}, {Guyon}, {Marois}, {Groff}, {Lozi}, {Vievard}, {Sahoo}, {Deo},
  {Jovanovic}, {Martinache}, {Wagner}, {Dupuy}, {Wahl}, {Letawsky}, {Li},
  {Zeng}, {Brandt}, {Michalik}, {Grady}, {Janson}, {Knapp}, {Kwon}, {Lawson},
  {McElwain}, {Uyama}, {Wisniewski}, \& {Tamura}}]{Currie2020b}
{Currie}, T., {Brandt}, T.~D., {Kuzuhara}, M., {et~al.} 2020{\natexlab{b}},
  \apjl, 904, L25, \dodoi{10.3847/2041-8213/abc631}

\bibitem[{{Duch{\^e}ne} {et~al.}(2020){Duch{\^e}ne}, {Rice}, {Hom}, {Zalesky},
  {Esposito}, {Millar-Blanchaer}, {Ren}, {Kalas}, {Fitzgerald}, {Arriaga},
  {Bruzzone}, {Bulger}, {Chen}, {Chiang}, {Cotten}, {Czekala}, {De Rosa},
  {Dong}, {Draper}, {Follette}, {Graham}, {Hung}, {Lopez}, {Macintosh},
  {Matthews}, {Mazoyer}, {Metchev}, {Patience}, {Perrin}, {Rameau}, {Song},
  {Stahl}, {Wang}, {Wolff}, {Zuckerman}, {Ammons}, {Bailey}, {Barman},
  {Chilcote}, {Doyon}, {Gerard}, {Goodsell}, {Greenbaum}, {Hibon}, {Ingraham},
  {Konopacky}, {Maire}, {Marchis}, {Marley}, {Marois}, {Nielsen},
  {Oppenheimer}, {Palmer}, {Poyneer}, {Pueyo}, {Rajan}, {Rantakyr{\"o}},
  {Ruffio}, {Savransky}, {Schneider}, {Sivaramakrishnan}, {Soummer}, {Thomas},
  \& {Ward-Duong}}]{Duchene2020}
{Duch{\^e}ne}, G., {Rice}, M., {Hom}, J., {et~al.} 2020, \aj, 159, 251,
  \dodoi{10.3847/1538-3881/ab8881}

\bibitem[{{Engler} {et~al.}(2019){Engler}, {Boccaletti}, {Schmid}, {Milli},
  {Augereau}, {Mazoyer}, {Maire}, {Henning}, {Avenhaus}, {Baudoz}, {Feldt},
  {Galicher}, {Hinkley}, {Lagrange}, {Mawet}, {Olofsson}, {Pantin}, {Perrot},
  \& {Stapelfeldt}}]{Engler2019}
{Engler}, N., {Boccaletti}, A., {Schmid}, H.~M., {et~al.} 2019, \aap, 622,
  A192, \dodoi{10.1051/0004-6361/201833542}

\bibitem[{{Esposito} {et~al.}(2020){Esposito}, {Kalas}, {Fitzgerald},
  {Millar-Blanchaer}, {Duch{\^e}ne}, {Patience}, {Hom}, {Perrin}, {De Rosa},
  {Chiang}, {Czekala}, {Macintosh}, {Graham}, {Ansdell}, {Arriaga}, {Bruzzone},
  {Bulger}, {Chen}, {Cotten}, {Dong}, {Draper}, {Follette}, {Hung}, {Lopez},
  {Matthews}, {Mazoyer}, {Metchev}, {Rameau}, {Ren}, {Rice}, {Song}, {Stahl},
  {Wang}, {Wolff}, {Zuckerman}, {Ammons}, {Bailey}, {Barman}, {Chilcote},
  {Doyon}, {Gerard}, {Goodsell}, {Greenbaum}, {Hibon}, {Hinkley}, {Ingraham},
  {Konopacky}, {Maire}, {Marchis}, {Marley}, {Marois}, {Nielsen},
  {Oppenheimer}, {Palmer}, {Poyneer}, {Pueyo}, {Rajan}, {Rantakyr{\"o}},
  {Ruffio}, {Savransky}, {Schneider}, {Sivaramakrishnan}, {Soummer}, {Thomas},
  \& {Ward-Duong}}]{Esposito2020}
{Esposito}, T.~M., {Kalas}, P., {Fitzgerald}, M.~P., {et~al.} 2020, \aj, 160,
  24, \dodoi{10.3847/1538-3881/ab9199}

\bibitem[{{Fitzgerald} {et~al.}(2007){Fitzgerald}, {Kalas}, {Duch{\^e}ne},
  {Pinte}, \& {Graham}}]{Fitzgerald2007}
{Fitzgerald}, M.~P., {Kalas}, P.~G., {Duch{\^e}ne}, G., {Pinte}, C., \&
  {Graham}, J.~R. 2007, \apj, 670, 536, \dodoi{10.1086/521344}

\bibitem[{{Gagn{\'e}} {et~al.}(2018){Gagn{\'e}}, {Mamajek}, {Malo}, {Riedel},
  {Rodriguez}, {Lafreni{\`e}re}, {Faherty}, {Roy-Loubier}, {Pueyo}, {Robin}, \&
  {Doyon}}]{Gagne2018}
{Gagn{\'e}}, J., {Mamajek}, E.~E., {Malo}, L., {et~al.} 2018, \apj, 856, 23,
  \dodoi{10.3847/1538-4357/aaae09}

\bibitem[{{Gaia Collaboration} {et~al.}(2018){Gaia Collaboration}, {Brown},
  {Vallenari}, {Prusti}, {de Bruijne}, {Babusiaux}, {Bailer-Jones}, {Biermann},
  {Evans}, {Eyer}, {Jansen}, {Jordi}, {Klioner}, {Lammers}, {Lindegren},
  {Luri}, {Mignard}, {Panem}, {Pourbaix}, {Randich}, {Sartoretti}, {Siddiqui},
  {Soubiran}, {van Leeuwen}, {Walton}, {Arenou}, {Bastian}, {Cropper},
  {Drimmel}, {Katz}, {Lattanzi}, {Bakker}, {Cacciari}, {Casta{\~n}eda},
  {Chaoul}, {Cheek}, {De Angeli}, {Fabricius}, {Guerra}, {Holl}, {Masana},
  {Messineo}, {Mowlavi}, {Nienartowicz}, {Panuzzo}, {Portell}, {Riello},
  {Seabroke}, {Tanga}, {Th{\'e}venin}, {Gracia-Abril}, {Comoretto},
  {Garcia-Reinaldos}, {Teyssier}, {Altmann}, {Andrae}, {Audard},
  {Bellas-Velidis}, {Benson}, {Berthier}, {Blomme}, {Burgess}, {Busso},
  {Carry}, {Cellino}, {Clementini}, {Clotet}, {Creevey}, {Davidson}, {De
  Ridder}, {Delchambre}, {Dell'Oro}, {Ducourant},
  {Fern{\'a}ndez-Hern{\'a}ndez}, {Fouesneau}, {Fr{\'e}mat}, {Galluccio},
  {Garc{\'\i}a-Torres}, {Gonz{\'a}lez-N{\'u}{\~n}ez}, {Gonz{\'a}lez-Vidal},
  {Gosset}, {Guy}, {Halbwachs}, {Hambly}, {Harrison}, {Hern{\'a}ndez},
  {Hestroffer}, {Hodgkin}, {Hutton}, {Jasniewicz}, {Jean-Antoine-Piccolo},
  {Jordan}, {Korn}, {Krone-Martins}, {Lanzafame}, {Lebzelter}, {L{\"o}ffler},
  {Manteiga}, {Marrese}, {Mart{\'\i}n-Fleitas}, {Moitinho}, {Mora}, {Muinonen},
  {Osinde}, {Pancino}, {Pauwels}, {Petit}, {Recio-Blanco}, {Richards},
  {Rimoldini}, {Robin}, {Sarro}, {Siopis}, {Smith}, {Sozzetti}, {S{\"u}veges},
  {Torra}, {van Reeven}, {Abbas}, {Abreu Aramburu}, {Accart}, {Aerts},
  {Altavilla}, {{\'A}lvarez}, {Alvarez}, {Alves}, {Anderson}, {Andrei},
  {Anglada Varela}, {Antiche}, {Antoja}, {Arcay}, {Astraatmadja}, {Bach},
  {Baker}, {Balaguer-N{\'u}{\~n}ez}, {Balm}, {Barache}, {Barata}, {Barbato},
  {Barblan}, {Barklem}, {Barrado}, {Barros}, {Barstow}, {Bartholom{\'e}
  Mu{\~n}oz}, {Bassilana}, {Becciani}, {Bellazzini}, {Berihuete}, {Bertone},
  {Bianchi}, {Bienaym{\'e}}, {Blanco-Cuaresma}, {Boch}, {Boeche}, {Bombrun},
  {Borrachero}, {Bossini}, {Bouquillon}, {Bourda}, {Bragaglia}, {Bramante},
  {Breddels}, {Bressan}, {Brouillet}, {Br{\"u}semeister}, {Brugaletta},
  {Bucciarelli}, {Burlacu}, {Busonero}, {Butkevich}, {Buzzi}, {Caffau},
  {Cancelliere}, {Cannizzaro}, {Cantat-Gaudin}, {Carballo}, {Carlucci},
  {Carrasco}, {Casamiquela}, {Castellani}, {Castro-Ginard}, {Charlot},
  {Chemin}, {Chiavassa}, {Cocozza}, {Costigan}, {Cowell}, {Crifo}, {Crosta},
  {Crowley}, {Cuypers}, {Dafonte}, {Damerdji}, {Dapergolas}, {David}, {David},
  {de Laverny}, {De Luise}, {De March}, {de Martino}, {de Souza}, {de Torres},
  {Debosscher}, {del Pozo}, {Delbo}, {Delgado}, {Delgado}, {Di Matteo},
  {Diakite}, {Diener}, {Distefano}, {Dolding}, {Drazinos}, {Dur{\'a}n},
  {Edvardsson}, {Enke}, {Eriksson}, {Esquej}, {Eynard Bontemps}, {Fabre},
  {Fabrizio}, {Faigler}, {Falc{\~a}o}, {Farr{\`a}s Casas}, {Federici},
  {Fedorets}, {Fernique}, {Figueras}, {Filippi}, {Findeisen}, {Fonti},
  {Fraile}, {Fraser}, {Fr{\'e}zouls}, {Gai}, {Galleti}, {Garabato},
  {Garc{\'\i}a-Sedano}, {Garofalo}, {Garralda}, {Gavel}, {Gavras}, {Gerssen},
  {Geyer}, {Giacobbe}, {Gilmore}, {Girona}, {Giuffrida}, {Glass}, {Gomes},
  {Granvik}, {Gueguen}, {Guerrier}, {Guiraud}, {Guti{\'e}rrez-S{\'a}nchez},
  {Haigron}, {Hatzidimitriou}, {Hauser}, {Haywood}, {Heiter}, {Helmi}, {Heu},
  {Hilger}, {Hobbs}, {Hofmann}, {Holland}, {Huckle}, {Hypki}, {Icardi},
  {Jan{\ss}en}, {Jevardat de Fombelle}, {Jonker}, {Juh{\'a}sz}, {Julbe},
  {Karampelas}, {Kewley}, {Klar}, {Kochoska}, {Kohley}, {Kolenberg},
  {Kontizas}, {Kontizas}, {Koposov}, {Kordopatis}, {Kostrzewa-Rutkowska},
  {Koubsky}, {Lambert}, {Lanza}, {Lasne}, {Lavigne}, {Le Fustec}, {Le
  Poncin-Lafitte}, {Lebreton}, {Leccia}, {Leclerc}, {Lecoeur-Taibi},
  {Lenhardt}, {Leroux}, {Liao}, {Licata}, {Lindstr{\o}m}, {Lister}, {Livanou},
  {Lobel}, {L{\'o}pez}, {Managau}, {Mann}, {Mantelet}, {Marchal}, {Marchant},
  {Marconi}, {Marinoni}, {Marschalk{\'o}}, {Marshall}, {Martino}, {Marton},
  {Mary}, {Massari}, {Matijevi{\v{c}}}, {Mazeh}, {McMillan}, {Messina},
  {Michalik}, {Millar}, {Molina}, {Molinaro}, {Moln{\'a}r}, {Montegriffo},
  {Mor}, {Morbidelli}, {Morel}, {Morris}, {Mulone}, {Muraveva}, {Musella},
  {Nelemans}, {Nicastro}, {Noval}, {O'Mullane}, {Ord{\'e}novic},
  {Ord{\'o}{\~n}ez-Blanco}, {Osborne}, {Pagani}, {Pagano}, {Pailler},
  {Palacin}, {Palaversa}, {Panahi}, {Pawlak}, {Piersimoni}, {Pineau}, {Plachy},
  {Plum}, {Poggio}, {Poujoulet}, {Pr{\v{s}}a}, {Pulone}, {Racero}, {Ragaini},
  {Rambaux}, {Ramos-Lerate}, {Regibo}, {Reyl{\'e}}, {Riclet}, {Ripepi}, {Riva},
  {Rivard}, {Rixon}, {Roegiers}, {Roelens}, {Romero-G{\'o}mez}, {Rowell},
  {Royer}, {Ruiz-Dern}, {Sadowski}, {Sagrist{\`a} Sell{\'e}s}, {Sahlmann},
  {Salgado}, {Salguero}, {Sanna}, {Santana-Ros}, {Sarasso}, {Savietto},
  {Schultheis}, {Sciacca}, {Segol}, {Segovia}, {S{\'e}gransan}, {Shih},
  {Siltala}, {Silva}, {Smart}, {Smith}, {Solano}, {Solitro}, {Sordo}, {Soria
  Nieto}, {Souchay}, {Spagna}, {Spoto}, {Stampa}, {Steele},
  {Steidelm{\"u}ller}, {Stephenson}, {Stoev}, {Suess}, {Surdej}, {Szabados},
  {Szegedi-Elek}, {Tapiador}, {Taris}, {Tauran}, {Taylor}, {Teixeira},
  {Terrett}, {Teyssand ier}, {Thuillot}, {Titarenko}, {Torra Clotet}, {Turon},
  {Ulla}, {Utrilla}, {Uzzi}, {Vaillant}, {Valentini}, {Valette}, {van Elteren},
  {Van Hemelryck}, {van Leeuwen}, {Vaschetto}, {Vecchiato}, {Veljanoski},
  {Viala}, {Vicente}, {Vogt}, {von Essen}, {Voss}, {Votruba}, {Voutsinas},
  {Walmsley}, {Weiler}, {Wertz}, {Wevers}, {Wyrzykowski}, {Yoldas},
  {{\v{Z}}erjal}, {Ziaeepour}, {Zorec}, {Zschocke}, {Zucker}, {Zurbach}, \&
  {Zwitter}}]{Gaia2018}
{Gaia Collaboration}, {Brown}, A.~G.~A., {Vallenari}, A., {et~al.} 2018, \aap,
  616, A1, \dodoi{10.1051/0004-6361/201833051}

\bibitem[{{Gibbs} {et~al.}(2019){Gibbs}, {Wagner}, {Apai}, {Mo{\'o}r},
  {Currie}, {Bonnefoy}, {Langlois}, \& {Lisse}}]{Gibbs2019}
{Gibbs}, A., {Wagner}, K., {Apai}, D., {et~al.} 2019, \aj, 157, 39,
  \dodoi{10.3847/1538-3881/aaf1bd}

\bibitem[{{Goebel} {et~al.}(2018){Goebel}, {Currie}, {Guyon}, {Brand t},
  {Groff}, {Jovanovic}, {Kasdin}, {Lozi}, {Hodapp}, {Martinache}, {Grady},
  {Hayashi}, {Kwon}, {McElwain}, {Yang}, \& {Tamura}}]{Goebel2018}
{Goebel}, S., {Currie}, T., {Guyon}, O., {et~al.} 2018, \aj, 156, 279,
  \dodoi{10.3847/1538-3881/aaeb24}

\bibitem[{{Groff} {et~al.}(2016){Groff}, {Chilcote}, {Kasdin}, {Galvin},
  {Loomis}, {Carr}, {Brand t}, {Knapp}, {Limbach}, {Guyon}, {Jovanovic},
  {McElwain}, {Takato}, \& {Hayashi}}]{Groff2016}
{Groff}, T.~D., {Chilcote}, J., {Kasdin}, N.~J., {et~al.} 2016, in Society of
  Photo-Optical Instrumentation Engineers (SPIE) Conference Series, Vol. 9908,
  Ground-based and Airborne Instrumentation for Astronomy VI, 99080O,
  \dodoi{10.1117/12.2233447}

\bibitem[{{Henyey} \& {Greenstein}(1941)}]{Henyey1941}
{Henyey}, L.~G., \& {Greenstein}, J.~L. 1941, \apj, 93, 70,
  \dodoi{10.1086/144246}

\bibitem[{{Hong}(1985)}]{Hong1985}
{Hong}, S.~S. 1985, \aap, 146, 67

\bibitem[{{Hughes} {et~al.}(2018){Hughes}, {Duch{\^e}ne}, \&
  {Matthews}}]{Hughes2018}
{Hughes}, A.~M., {Duch{\^e}ne}, G., \& {Matthews}, B.~C. 2018, \araa, 56, 541,
  \dodoi{10.1146/annurev-astro-081817-052035}

\bibitem[{{Jovanovic} {et~al.}(2015){Jovanovic}, {Martinache}, {Guyon},
  {Clergeon}, {Singh}, {Kudo}, {Garrel}, {Newman}, {Doughty}, {Lozi}, {Males},
  {Minowa}, {Hayano}, {Takato}, {Morino}, {Kuhn}, {Serabyn}, {Norris},
  {Tuthill}, {Schworer}, {Stewart}, {Close}, {Huby}, {Perrin}, {Lacour},
  {Gauchet}, {Vievard}, {Murakami}, {Oshiyama}, {Baba}, {Matsuo}, {Nishikawa},
  {Tamura}, {Lai}, {Marchis}, {Duchene}, {Kotani}, \&
  {Woillez}}]{Jovanovic2015}
{Jovanovic}, N., {Martinache}, F., {Guyon}, O., {et~al.} 2015, \pasp, 127, 890,
  \dodoi{10.1086/682989}

\bibitem[{{Kalas} {et~al.}(2005){Kalas}, {Graham}, \& {Clampin}}]{Kalas2005}
{Kalas}, P., {Graham}, J.~R., \& {Clampin}, M. 2005, \nat, 435, 1067,
  \dodoi{10.1038/nature03601}

\bibitem[{{Kenyon} \& {Bromley}(2008)}]{Kenyon2008}
{Kenyon}, S.~J., \& {Bromley}, B.~C. 2008, \apjs, 179, 451,
  \dodoi{10.1086/591794}

\bibitem[{{Lafreni{\`e}re} {et~al.}(2007){Lafreni{\`e}re}, {Marois}, {Doyon},
  {Nadeau}, \& {Artigau}}]{Lafreniere2007}
{Lafreni{\`e}re}, D., {Marois}, C., {Doyon}, R., {Nadeau}, D., \& {Artigau},
  {\'E}. 2007, \apj, 660, 770, \dodoi{10.1086/513180}

\bibitem[{{Lagrange} {et~al.}(2010){Lagrange}, {Bonnefoy}, {Chauvin}, {Apai},
  {Ehrenreich}, {Boccaletti}, {Gratadour}, {Rouan}, {Mouillet}, {Lacour}, \&
  {Kasper}}]{Lagrange2010}
{Lagrange}, A.-M., {Bonnefoy}, M., {Chauvin}, G., {et~al.} 2010, Science, 329,
  57, \dodoi{10.1126/science.1187187}

\bibitem[{{Lawson} {et~al.}(2020){Lawson}, {Currie}, {Wisniewski}, {Tamura},
  {Schneider}, {Augereau}, {Brandt}, {Guyon}, {Kasdin}, {Groff}, {Lozi},
  {Chilcote}, {Hodapp}, {Jovanovic}, {Martinache}, {Skaf}, {Akiyama},
  {Henning}, {Knapp}, {Kwon}, {Mayama}, {McElwain}, {Sitko}, {Asensio-Torres},
  {Uyama}, \& {Wagner}}]{Lawson2020}
{Lawson}, K., {Currie}, T., {Wisniewski}, J.~P., {et~al.} 2020, \aj, 160, 163,
  \dodoi{10.3847/1538-3881/ababa6}

\bibitem[{{Lisse} {et~al.}(2017){Lisse}, {Sitko}, {Russell}, {Marengo},
  {Currie}, {Melis}, {Mittal}, \& {Song}}]{Lisse2017}
{Lisse}, C.~M., {Sitko}, M.~L., {Russell}, R.~W., {et~al.} 2017, \apjl, 840,
  L20, \dodoi{10.3847/2041-8213/aa6ea3}

\bibitem[{{Lozi} {et~al.}(2018){Lozi}, {Guyon}, {Jovanovic}, {Goebel},
  {Pathak}, {Skaf}, {Sahoo}, {Norris}, {Martinache}, {N'Diaye}, {Mazin},
  {Walter}, {Tuthill}, {Kudo}, {Kawahara}, {Kotani}, {Ireland}, {Cvetojevic},
  {Huby}, {Lacour}, {Vievard}, {Groff}, {Chilcote}, {Kasdin}, {Knight}, {Snik},
  {Doelman}, {Minowa}, {Clergeon}, {Takato}, {Tamura}, {Currie}, {Takami}, \&
  {Hayashi}}]{Lozi2018}
{Lozi}, J., {Guyon}, O., {Jovanovic}, N., {et~al.} 2018, in Society of
  Photo-Optical Instrumentation Engineers (SPIE) Conference Series, Vol. 10703,
  \procspie, 1070359, \dodoi{10.1117/12.2314282}

\bibitem[{{Macintosh} {et~al.}(2015){Macintosh}, {Graham}, {Barman}, {De Rosa},
  {Konopacky}, {Marley}, {Marois}, {Nielsen}, {Pueyo}, {Rajan}, {Rameau},
  {Saumon}, {Wang}, {Patience}, {Ammons}, {Arriaga}, {Artigau}, {Beckwith},
  {Brewster}, {Bruzzone}, {Bulger}, {Burningham}, {Burrows}, {Chen}, {Chiang},
  {Chilcote}, {Dawson}, {Dong}, {Doyon}, {Draper}, {Duch{\^e}ne}, {Esposito},
  {Fabrycky}, {Fitzgerald}, {Follette}, {Fortney}, {Gerard}, {Goodsell},
  {Greenbaum}, {Hibon}, {Hinkley}, {Cotten}, {Hung}, {Ingraham},
  {Johnson-Groh}, {Kalas}, {Lafreniere}, {Larkin}, {Lee}, {Line}, {Long},
  {Maire}, {Marchis}, {Matthews}, {Max}, {Metchev}, {Millar-Blanchaer},
  {Mittal}, {Morley}, {Morzinski}, {Murray-Clay}, {Oppenheimer}, {Palmer},
  {Patel}, {Perrin}, {Poyneer}, {Rafikov}, {Rantakyr{\"o}}, {Rice}, {Rojo},
  {Rudy}, {Ruffio}, {Ruiz}, {Sadakuni}, {Saddlemyer}, {Salama}, {Savransky},
  {Schneider}, {Sivaramakrishnan}, {Song}, {Soummer}, {Thomas}, {Vasisht},
  {Wallace}, {Ward-Duong}, {Wiktorowicz}, {Wolff}, \&
  {Zuckerman}}]{Macintosh2015}
{Macintosh}, B., {Graham}, J.~R., {Barman}, T., {et~al.} 2015, Science, 350,
  64, \dodoi{10.1126/science.aac5891}

\bibitem[{{Marois} {et~al.}(2006){Marois}, {Lafreni{\`e}re}, {Doyon},
  {Macintosh}, \& {Nadeau}}]{Marois2006}
{Marois}, C., {Lafreni{\`e}re}, D., {Doyon}, R., {Macintosh}, B., \& {Nadeau},
  D. 2006, \apj, 641, 556, \dodoi{10.1086/500401}

\bibitem[{{Marois} {et~al.}(2010){Marois}, {Macintosh}, \&
  {V{\'e}ran}}]{Marois2010b}
{Marois}, C., {Macintosh}, B., \& {V{\'e}ran}, J.-P. 2010, in Society of
  Photo-Optical Instrumentation Engineers (SPIE) Conference Series, Vol. 7736,
  \procspie, 77361J, \dodoi{10.1117/12.857225}

\bibitem[{{Mawet} {et~al.}(2014){Mawet}, {Milli}, {Wahhaj}, {Pelat}, {Absil},
  {Delacroix}, {Boccaletti}, {Kasper}, {Kenworthy}, {Marois}, {Mennesson}, \&
  {Pueyo}}]{Mawet2014}
{Mawet}, D., {Milli}, J., {Wahhaj}, Z., {et~al.} 2014, \apj, 792, 97,
  \dodoi{10.1088/0004-637X/792/2/97}

\bibitem[{{Millar-Blanchaer} {et~al.}(2015){Millar-Blanchaer}, {Graham},
  {Pueyo}, {Kalas}, {Dawson}, {Wang}, {Perrin}, {moon}, {Macintosh}, {Ammons},
  {Barman}, {Cardwell}, {Chen}, {Chiang}, {Chilcote}, {Cotten}, {De Rosa},
  {Draper}, {Dunn}, {Duch{\^e}ne}, {Esposito}, {Fitzgerald}, {Follette},
  {Goodsell}, {Greenbaum}, {Hartung}, {Hibon}, {Hinkley}, {Ingraham},
  {Jensen-Clem}, {Konopacky}, {Larkin}, {Long}, {Maire}, {Marchis}, {Marley},
  {Marois}, {Morzinski}, {Nielsen}, {Palmer}, {Oppenheimer}, {Poyneer},
  {Rajan}, {Rantakyr{\"o}}, {Ruffio}, {Sadakuni}, {Saddlemyer}, {Schneider},
  {Sivaramakrishnan}, {Soummer}, {Thomas}, {Vasisht}, {Vega}, {Wallace},
  {Ward-Duong}, {Wiktorowicz}, \& {Wolff}}]{Millar-Blanchaer2015}
{Millar-Blanchaer}, M.~A., {Graham}, J.~R., {Pueyo}, L., {et~al.} 2015, \apj,
  811, 18, \dodoi{10.1088/0004-637X/811/1/18}

\bibitem[{{Milli} {et~al.}(2017){Milli}, {Vigan}, {Mouillet}, {Lagrange},
  {Augereau}, {Pinte}, {Mawet}, {Schmid}, {Boccaletti}, {Matr{\`a}}, {Kral},
  {Ertel}, {Chauvin}, {Bazzon}, {M{\'e}nard}, {Beuzit}, {Thalmann}, {Dominik},
  {Feldt}, {Henning}, {Min}, {Girard}, {Galicher}, {Bonnefoy}, {Fusco}, {de
  Boer}, {Janson}, {Maire}, {Mesa}, {Schlieder}, \& {Sphere
  Consortium}}]{Milli2017}
{Milli}, J., {Vigan}, A., {Mouillet}, D., {et~al.} 2017, \aap, 599, A108,
  \dodoi{10.1051/0004-6361/201527838}

\bibitem[{{Pueyo}(2016)}]{Pueyo2016}
{Pueyo}, L. 2016, \apj, 824, 117, \dodoi{10.3847/0004-637X/824/2/117}

\bibitem[{{Ribas} {et~al.}(2015){Ribas}, {Bouy}, \& {Mer{\'\i}n}}]{Ribas2015}
{Ribas}, {\'A}., {Bouy}, H., \& {Mer{\'\i}n}, B. 2015, \aap, 576, A52,
  \dodoi{10.1051/0004-6361/201424846}

\bibitem[{{Schneider} {et~al.}(2014){Schneider}, {Grady}, {Hines}, {Stark},
  {Debes}, {Carson}, {Kuchner}, {Perrin}, {Weinberger}, {Wisniewski},
  {Silverstone}, {Jang-Condell}, {Henning}, {Woodgate}, {Serabyn},
  {Moro-Martin}, {Tamura}, {Hinz}, \& {Rodigas}}]{Schneider2014}
{Schneider}, G., {Grady}, C.~A., {Hines}, D.~C., {et~al.} 2014, \aj, 148, 59,
  \dodoi{10.1088/0004-6256/148/4/59}

\bibitem[{{Smith} \& {Terrile}(1984)}]{Smith1984}
{Smith}, B.~A., \& {Terrile}, R.~J. 1984, Science, 226, 1421,
  \dodoi{10.1126/science.226.4681.1421}

\bibitem[{{Soummer} {et~al.}(2012){Soummer}, {Pueyo}, \&
  {Larkin}}]{Soummer2012}
{Soummer}, R., {Pueyo}, L., \& {Larkin}, J. 2012, \apjl, 755, L28,
  \dodoi{10.1088/2041-8205/755/2/L28}

\bibitem[{{Sparks} \& {Ford}(2002)}]{SparksFord2002}
{Sparks}, W.~B., \& {Ford}, H.~C. 2002, \apj, 578, 543, \dodoi{10.1086/342401}

\bibitem[{{Spiegel} \& {Burrows}(2012)}]{Spiegel2012}
{Spiegel}, D.~S., \& {Burrows}, A. 2012, \apj, 745, 174,
  \dodoi{10.1088/0004-637X/745/2/174}

\bibitem[{{Steiger} {et~al.}(2021){Steiger}, {Currie}, {Brandt}, {Guyon},
  {Kuzuhara}, {Chilcote}, {Groff}, {Lozi}, {Walter}, {Fruitwala}, {Bailey},
  {Zobrist}, {Swimmer}, {Lipartito}, {Smith}, {Bockstiegel}, {Meeker},
  {Coiffard}, {Dodkins}, {Szypryt}, {Davis}, {Daal}, {Bumble}, {Vievard},
  {Sahoo}, {Deo}, {Jovanovic}, {Martinache}, {Doppmann}, {Tamura}, {Kasdin}, \&
  {Mazin}}]{Steiger2021}
{Steiger}, S., {Currie}, T., {Brandt}, T.~D., {et~al.} 2021, \aj, 162, 44,
  \dodoi{10.3847/1538-3881/ac02cc}

\bibitem[{{Thalmann} {et~al.}(2011){Thalmann}, {Janson}, {Buenzli}, {Brandt},
  {Wisniewski}, {Moro-Mart{\'\i}n}, {Usuda}, {Schneider}, {Carson}, {McElwain},
  {Grady}, {Goto}, {Abe}, {Brandner}, {Dominik}, {Egner}, {Feldt}, {Fukue},
  {Golota}, {Guyon}, {Hashimoto}, {Hayano}, {Hayashi}, {Hayashi}, {Henning},
  {Hodapp}, {Ishii}, {Iye}, {Kandori}, {Knapp}, {Kudo}, {Kusakabe}, {Kuzuhara},
  {Matsuo}, {Miyama}, {Morino}, {Nishimura}, {Pyo}, {Serabyn}, {Suto},
  {Suzuki}, {Takahashi}, {Takami}, {Takato}, {Terada}, {Tomono}, {Turner},
  {Watanabe}, {Yamada}, {Takami}, \& {Tamura}}]{Thalmann2011}
{Thalmann}, C., {Janson}, M., {Buenzli}, E., {et~al.} 2011, \apjl, 743, L6,
  \dodoi{10.1088/2041-8205/743/1/L6}

\bibitem[{{Thalmann} {et~al.}(2013){Thalmann}, {Janson}, {Buenzli}, {Brandt},
  {Wisniewski}, {Dominik}, {Carson}, {McElwain}, {Currie}, {Knapp},
  {Moro-Mart{\'\i}n}, {Usuda}, {Abe}, {Brandner}, {Egner}, {Feldt}, {Golota},
  {Goto}, {Guyon}, {Hashimoto}, {Hayano}, {Hayashi}, {Hayashi}, {Henning},
  {Hodapp}, {Ishii}, {Iye}, {Kandori}, {Kudo}, {Kusakabe}, {Kuzuhara}, {Kwon},
  {Matsuo}, {Mayama}, {Miyama}, {Morino}, {Nishimura}, {Pyo}, {Serabyn},
  {Suto}, {Suzuki}, {Takami}, {Takato}, {Terada}, {Tomono}, {Turner},
  {Watanabe}, {Yamada}, {Takami}, \& {Tamura}}]{Thalmann2013}
---. 2013, \apjl, 763, L29, \dodoi{10.1088/2041-8205/763/2/L29}

\bibitem[{{Wyatt}(2008)}]{Wyatt2008}
{Wyatt}, M.~C. 2008, \araa, 46, 339,
  \dodoi{10.1146/annurev.astro.45.051806.110525}

\end{thebibliography}
\bibliographystyle{aasjournal}
\end{document}